\documentclass[twocolumn,superscriptaddress,showkeys,nofootinbib]{revtex4-2}
\usepackage{graphicx}

\RequirePackage{lineno}
\usepackage{hyperref}
\usepackage{textgreek}
\usepackage{booktabs}
\usepackage{color,soul}
\usepackage{xspace}
\newcommand{\nuGeN} {{{$\mathrm{\nu}$GeN }}}
\newcommand{\CEvNS} {{CE{$\mathrm{\nu}$NS }}}

\begin{document}

\title{New constraints on coherent elastic neutrino–nucleus scattering by the \nuGeN experiment}
\thanks{This work is supported within the State Project ``Science'' by the Ministry of Science and Higher Education of the Russian Federation under Contract 075-15-2024-541.}

\author{ V.~Belov}
\affiliation{Dzhelepov Laboratory of Nuclear Problems, Joint Institute for Nuclear Research, 6 Joliot-Curie, 141980, Dubna, Russia}
\affiliation{Institute for Nuclear Research of the Russian Academy of Sciences, 7a Prospect 60-letiya Oktyabrya, 117312, Moscow, Russia}

\author{A.~Bystryakov}
\affiliation{Dzhelepov Laboratory of Nuclear Problems, Joint Institute for Nuclear Research, 6 Joliot-Curie, 141980, Dubna, Russia}
\affiliation{Lebedev Physical Institute of the Russian Academy of Sciences, 53 Leninskiy Prospect, 119991, Moscow, Russia}
\affiliation{Dubna State University, 19 Universitetskaya St., 141980, Dubna, Russia}

\author{ M.~Danilov}
\affiliation{Lebedev Physical Institute of the Russian Academy of Sciences, 53 Leninskiy Prospect, 119991, Moscow, Russia}
\affiliation{Institute for Nuclear Research of the Russian Academy of Sciences, 7a Prospect 60-letiya Oktyabrya, 117312, Moscow, Russia}

\author{S.~Evseev}
\affiliation{Dzhelepov Laboratory of Nuclear Problems, Joint Institute for Nuclear Research, 6 Joliot-Curie, 141980, Dubna, Russia}

\author{M.~Fomina}
\affiliation{Dzhelepov Laboratory of Nuclear Problems, Joint Institute for Nuclear Research, 6 Joliot-Curie, 141980, Dubna, Russia}

\author{G.~Ignatov}
\affiliation{Lebedev Physical Institute of the Russian Academy of Sciences, 53 Leninskiy Prospect, 119991, Moscow, Russia}
\affiliation{Moscow Institute of Physics and Technology, 9 Institutskiy per., 141700, Dolgoprudny, Russia}

\author{S.~Kazartsev}
\affiliation{Dzhelepov Laboratory of Nuclear Problems, Joint Institute for Nuclear Research, 6 Joliot-Curie, 141980, Dubna, Russia}
\affiliation{Institute for Nuclear Research of the Russian Academy of Sciences, 7a Prospect 60-letiya Oktyabrya, 117312, Moscow, Russia}

\author{ J.~Khushvaktov}
\affiliation{Dzhelepov Laboratory of Nuclear Problems, Joint Institute for Nuclear Research, 6 Joliot-Curie, 141980, Dubna, Russia}

\author{ T.~Khussainov}
\affiliation{Dzhelepov Laboratory of Nuclear Problems, Joint Institute for Nuclear Research, 6 Joliot-Curie, 141980, Dubna, Russia}
\affiliation{Institute of Nuclear Physics of the Ministry of Energy of the Republic of Kazakhstan, 1 Ibragimov Street, 050032, Almaty, Kazakhstan}

\author{ A.~Konovalov}
\affiliation{Lebedev Physical Institute of the Russian Academy of Sciences, 53 Leninskiy Prospect, 119991, Moscow, Russia}
\affiliation{Institute for Nuclear Research of the Russian Academy of Sciences, 7a Prospect 60-letiya Oktyabrya, 117312, Moscow, Russia}

\author{A.~Kuznetsov}
\affiliation{Dzhelepov Laboratory of Nuclear Problems, Joint Institute for Nuclear Research, 6 Joliot-Curie, 141980, Dubna, Russia}

\author{A.~Lubashevskiy}
\email[]{lubashev@jinr.ru}
\affiliation{Dzhelepov Laboratory of Nuclear Problems, Joint Institute for Nuclear Research, 6 Joliot-Curie, 141980, Dubna, Russia}
\affiliation{Lebedev Physical Institute of the Russian Academy of Sciences, 53 Leninskiy Prospect, 119991, Moscow, Russia}
\affiliation{Institute for Nuclear Research of the Russian Academy of Sciences, 7a Prospect 60-letiya Oktyabrya, 117312, Moscow, Russia}

\author{D.~Medvedev}
\affiliation{Dzhelepov Laboratory of Nuclear Problems, Joint Institute for Nuclear Research, 6 Joliot-Curie, 141980, Dubna, Russia}

\author{D.~Ponomarev}
\affiliation{Dzhelepov Laboratory of Nuclear Problems, Joint Institute for Nuclear Research, 6 Joliot-Curie, 141980, Dubna, Russia}
\affiliation{Lebedev Physical Institute of the Russian Academy of Sciences, 53 Leninskiy Prospect, 119991, Moscow, Russia}
\affiliation{Institute for Nuclear Research of the Russian Academy of Sciences, 7a Prospect 60-letiya Oktyabrya, 117312, Moscow, Russia}

\author{D.~Sautner}
\affiliation{Moscow Institute of Physics and Technology, 9 Institutskiy per., 141700, Dolgoprudny, Russia}

\author{K.~Shakhov}
\affiliation{Dzhelepov Laboratory of Nuclear Problems, Joint Institute for Nuclear Research, 6 Joliot-Curie, 141980, Dubna, Russia}

\author{ E.~Shevchik}
\affiliation{Dzhelepov Laboratory of Nuclear Problems, Joint Institute for Nuclear Research, 6 Joliot-Curie, 141980, Dubna, Russia}

\author{M.~Shirchenko}
\affiliation{Dzhelepov Laboratory of Nuclear Problems, Joint Institute for Nuclear Research, 6 Joliot-Curie, 141980, Dubna, Russia}
\affiliation{Institute for Nuclear Research of the Russian Academy of Sciences, 7a Prospect 60-letiya Oktyabrya, 117312, Moscow, Russia}

\author{S.~Rozov}
\affiliation{Dzhelepov Laboratory of Nuclear Problems, Joint Institute for Nuclear Research, 6 Joliot-Curie, 141980, Dubna, Russia}

\author{I.~Rozova}
\affiliation{Dzhelepov Laboratory of Nuclear Problems, Joint Institute for Nuclear Research, 6 Joliot-Curie, 141980, Dubna, Russia}

\author{S.~Vasilyev}
\affiliation{Dzhelepov Laboratory of Nuclear Problems, Joint Institute for Nuclear Research, 6 Joliot-Curie, 141980, Dubna, Russia}

\author{E.~Yakushev}
\affiliation{Dzhelepov Laboratory of Nuclear Problems, Joint Institute for Nuclear Research, 6 Joliot-Curie, 141980, Dubna, Russia}

\author{I.~Zhitnikov}
\affiliation{Dzhelepov Laboratory of Nuclear Problems, Joint Institute for Nuclear Research, 6 Joliot-Curie, 141980, Dubna, Russia}
\affiliation{Institute for Nuclear Research of the Russian Academy of Sciences, 7a Prospect 60-letiya Oktyabrya, 117312, Moscow, Russia}

\author{D.~Zinatulina}
\affiliation{Dzhelepov Laboratory of Nuclear Problems, Joint Institute for Nuclear Research, 6 Joliot-Curie, 141980, Dubna, Russia}

\collaboration{\nuGeN Collaboration}

\date{\today}

\begin{abstract}

The \nuGeN experiment searches for coherent elastic neutrino-nucleus scattering (CE{$\mathrm{\nu}$NS}) at the Kalinin Nuclear Power Plant. A 1.41-kg high-purity low-threshold germanium detector surrounded by active and passive shielding is deployed at the minimal distance of 11.1 m allowed by the lifting mechanism from the center of reactor core, utilizing one of the highest antineutrino fluxes among the competing experiments. The direct comparison of the count rates obtained during reactor-ON and reactor-OFF periods with the energy threshold of 0.29~keV$_{ee}$ shows no statistically significant difference. New upper limits on the number of \CEvNS events are evaluated on the basis of the residual ON$-$OFF count rate spectrum.

\end{abstract}


\maketitle

\section{Introduction}

Investigation of neutrino properties is one of the fast-developing areas of modern particle physics. Recently, significant advances have been achieved in the search for and study of coherent elastic neutrino–nucleus scattering, a process predicted within the Standard Model~\cite{Fre74},~\cite{Dru84}. Due to a small momentum transfer, the neutrino interacts simultaneously with all nucleons, and the cross-section of such a process is enhanced by several orders of magnitude in comparison with other neutrino interactions at the same energy~\cite{Akimov:2022oyb,Akimov:2019wtg}. The differential cross-section of \CEvNS for a spin-zero nucleus with a mass $M$ can be expressed as~\cite{And11}:
\begin{equation}
\Big(\frac{d\sigma}{dT}\Big) = \frac{G^{2}_{F}}{4 \pi} Q^2_W M \Big[1 - \frac{M T}{2E_{\nu}^2} \Big] F^2(Q^2),
\end{equation}
where $T$ is the nuclear recoil energy, $E_{\nu}$ is the neutrino energy, $Q$ is the transferred momentum, $F(Q^{2})$ is the nuclear form-factor. The Fermi constant is labelled as $G_{F}$, and $Q_W = N - (1-4 sin^2 \theta_W) Z$ is the weak charge of a nucleus. Since the predicted value of the Weinberg angle at low energies is $sin^2 \theta_W = 0.23867\pm0.00016$~\cite{angleW}, the full \CEvNS cross-section $\sigma$ is almost proportional to  $N^2$, the squared number of nuclear target neutrons. Studying \CEvNS allows us to test the Standard Model, search for non-standard neutrino interactions, probe some aspects of nuclear physics, and perform other investigations~\cite{Rev22}.

The first observation of \CEvNS~\cite{COHERENT} using neutrinos produced by the SNS accelerator~\cite{COHERENT:2021yvp} was reported by the COHERENT experiment. The search for coherent elastic scattering of reactor antineutrinos off nuclei is being actively performed these days. There are many experiments currently running or under construction (\cite{Bon21, Augier2023, Ang19, Ale19, Colaresi:2022obx, RED-100:2019rpf, Karmakar:2024ydi,NEON:2022hbk} and elsewhere). The sensitivity of these experiments is approaching the \CEvNS detection. The \CEvNS search at reactors is of particular interest due to the possibility of both looking for the physics beyond the Standard Model and monitoring reactor operation directly with a possibility to observe neutrino below the 1.8 MeV inverse beta decay energy threshold~\cite{RevModPhys.92.011003}. Due to a higher neutrino cross-section, the detectors typically have a much smaller size in comparison with those that are aimed to register reactor antineutrinos with the help of the inverse beta decay.

\section{Experimental setup}
\begin{figure}[hbt]
  \begin{center}
    \includegraphics[width=1\linewidth]{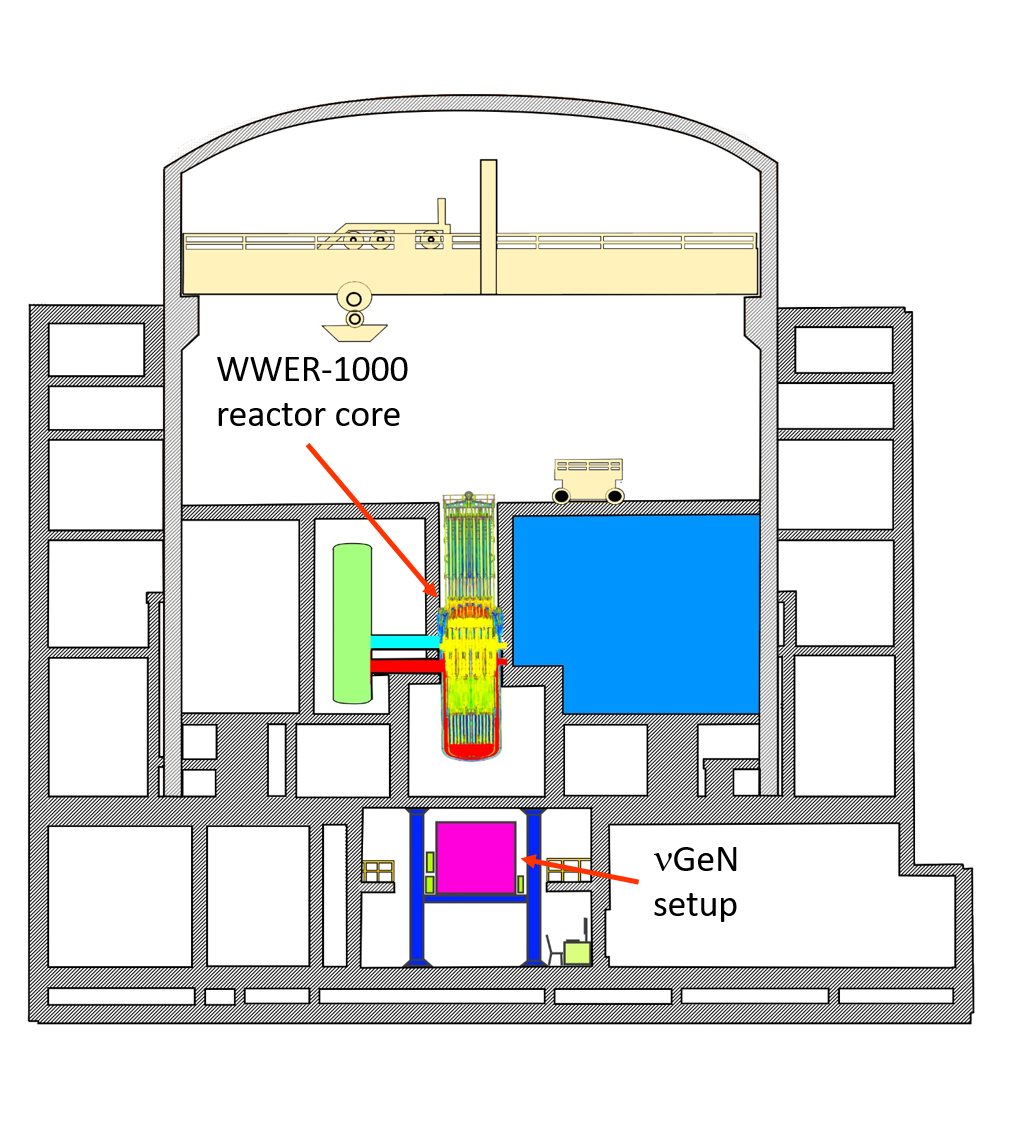}
    \caption{\label{fig:reactorsite} Scheme of the reactor unit and the site of the \nuGeN experimental setup (not to scale).}
  \end{center}
\end{figure}

A high neutrino flux, low background, large mass of the target, and low energy threshold are needed to detect CE{$\mathrm{\nu}$NS}. The signals of interest can be obscured by some backgrounds, like elastic neutron scattering or electronic noise, which complicates this task. The cosmogenic background is one of the most dangerous for \CEvNS experiments with a shallow overburden, and it is hard to mitigate due to the production of secondary fast neutrons inside the shielding of the setup~\cite{Bonet:2021wjw}. The \nuGeN experiment has one of the best locations to search for \CEvNS – at the Kalinin Nuclear Power Plant (KNPP) in Udomlya, Russia, Unit \#3, near the 3.1-GW$_{th}$ reactor of the WWER-1000 type~\cite{Bel15}. The close vicinity of the reactor core (11.1~m) allows utilizing one of the largest possible fluxes of antineutrinos of up to $\sim$4.4$\cdot10^{13}$~$\mathrm{cm^{-2} s^{-1}}$. The scheme of the reactor site and of the setup location (Room A-336~\cite{Ale22}) are shown in Figure~\ref{fig:reactorsite}.
The experimental site is located just under the reactor core, which together with other construction materials of the reactor provides shielding from cosmic rays of about 50~m~w.e. Such overburden allows us to completely remove the hadronic component of the cosmic flux~\cite{HAUSSER1993223}.
The muon flux suppression factor measured at a similar reactor unit of KNPP is up to 13 times, depending on the zenith angle~\cite{Ale16}. Therefore, the background from secondary neutrons is significantly reduced in comparison with the one at a shallow laboratory. The dimensions of the room are 8.7 by 9.3~m, with a height of 4.1~m. The experimental setup is located in the centre of the room on a lifting device, which allows changing the distance to the center of the reactor core from 12.5 to 11.1 m, with the corresponding change of the neutrino flux from about 3.4$\cdot10^{13}$ to 4.4$\cdot10^{13}$~$\mathrm{cm^{-2} s^{-1}}$ according to the calculation method from ~\cite{Bed07}.

\begin{figure}[hbt]
  \begin{center}
    \includegraphics[width=1\linewidth]{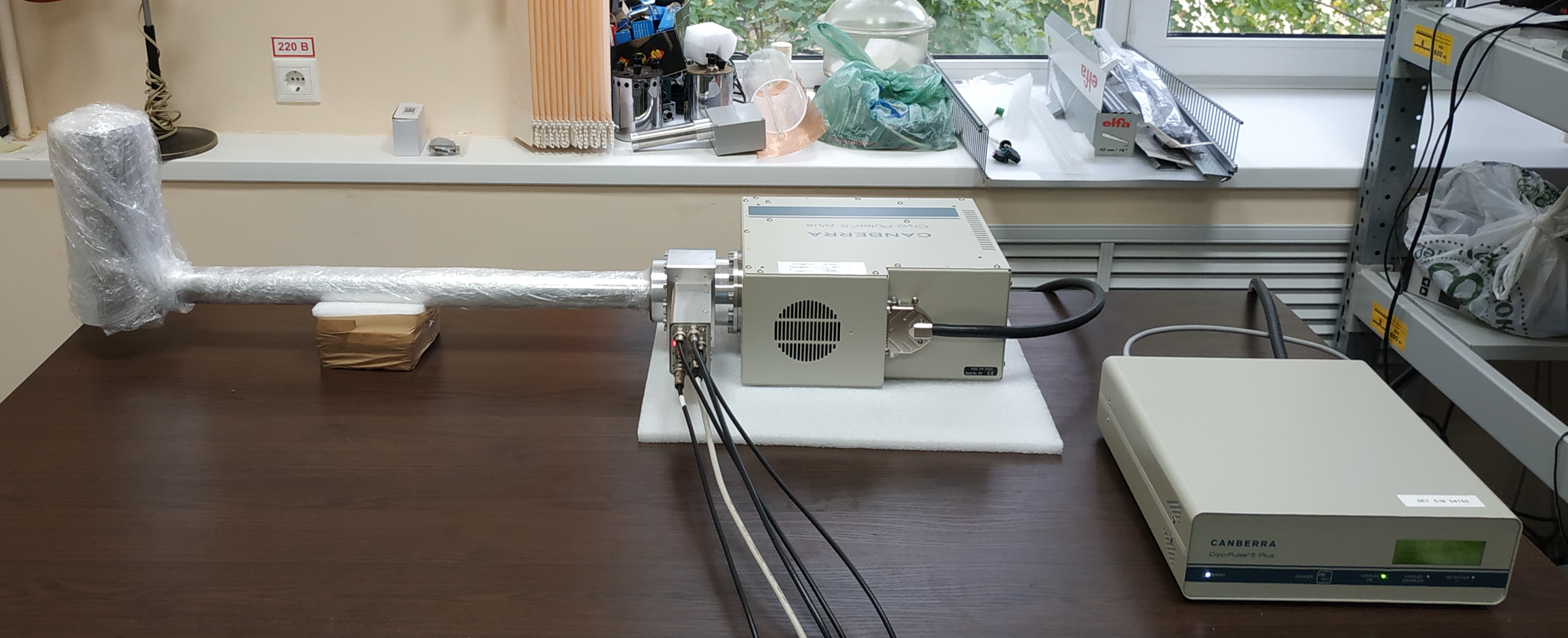}
    \caption{\label{fig:detector} The \nuGeN detector before deployment at KNPP.}
  \end{center}
\end{figure}

The \nuGeN experiment uses a custom-designed high-purity germanium (HPGe) detector manufactured by Mirion Technologies (Canberra Lingolsheim)~\cite{Mir20}. The p-type germanium crystal has a cylindrical shape with a diameter of 70 mm and a height of 70 mm. The sensitive volume of the detector is 265 cm$^3$, which corresponds to an active mass of 1.41~kg. The detector is installed inside the cryostat made mostly of low-background aluminum and copper. The photo of the germanium detector before installation is shown in Figure~\ref{fig:detector}. The germanium crystal is cooled by the electrically powered pulse tube cooler, model Cryo-Pulse 5 Plus (CP5+)~\cite{Cry20}. The cooling temperature of the detector is set to -185$^\circ$C. This temperature is optimal for the current detector in order to minimize the noise level and improve the energy resolution. The cooling power depends on the ambient temperature and is typically about 80~W. The detector is equipped with a 60-cm-long cooling rod allowing for sufficient shielding from external radiation.

\begin{figure}[hbt]
  \begin{center}
    \includegraphics[width=1\linewidth]{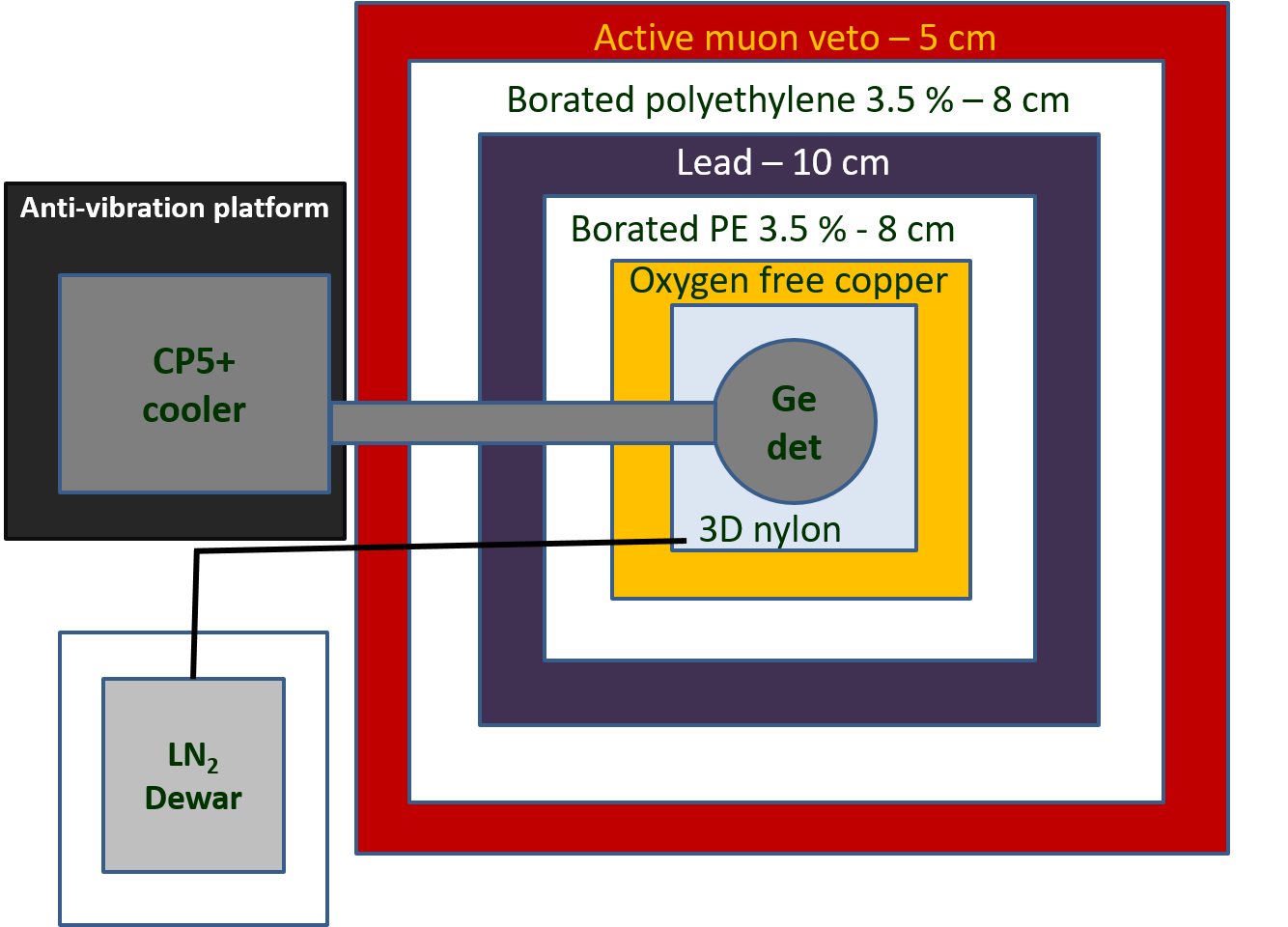}
    \caption{\label{fig:shielding} Scheme of the \nuGeN shielding (top view, not to scale).}
  \end{center}
\end{figure}

The innermost part of the shielding is made of 3D-printed nylon which displaces the air with a potential contamination of radon. The total density of the layer was set to 15\% of nylon density in order to make the layer soft enough to prevent any damage of the cryostat during the installation of the shielding. The thickness of the nylon layer is 16-46 mm, depending on the direction. The further layers are as follows~-- 10~cm of oxygen-free copper, 8~cm of borated (3.5\%) polyethylene, 10~cm of lead, then another 8~cm layer of borated polyethylene, and a 5-cm-thick active muon veto made of plastic scintillator panels. The radon level inside the shielding is further decreased by using nitrogen expulsion. The cryocooler of the detector is placed on an dynamic antivibration platform, TS-C30~\cite{Vib22}, in order to minimize vibrations from the surrounding equipment. The scheme of the setup structure including the passive and active shielding, is shown in Figure~\ref{fig:shielding}.

\begin{figure}[htb]
  \begin{center}
    \includegraphics[width=1\linewidth]{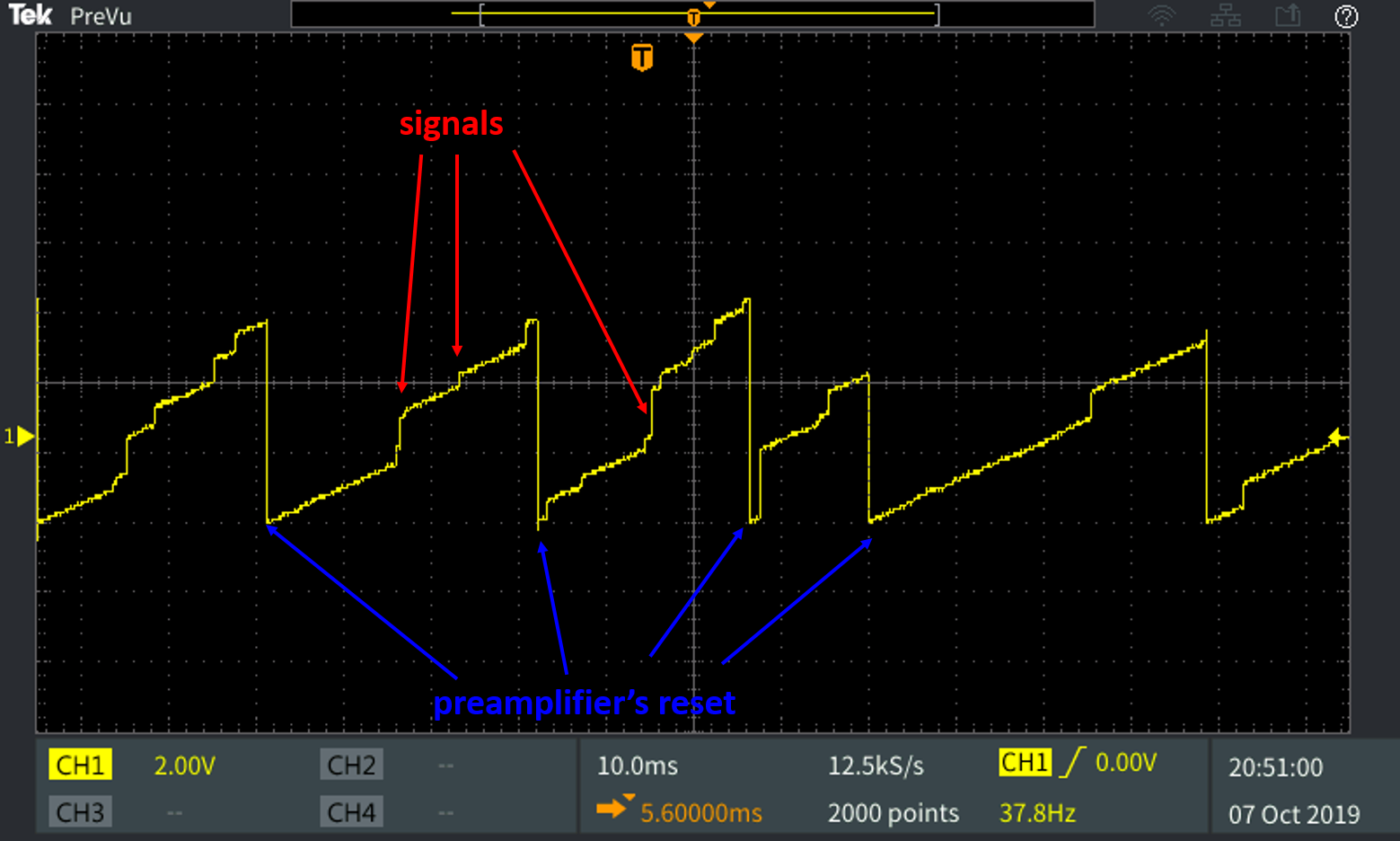}
    \caption{\label{fig:pulses} Screenshot from the digital oscilloscope demonstrates a typical output from the preamplifier of the HPGe detector under laboratory conditions without the described shielding.}
  \end{center}
\end{figure}

The germanium diode is electrically depleted by a positive operating voltage of 2300~V produced by a CAEN power supply, Mod.~No.~1471HA. Ionization energy losses induced by incoming particles passing through the HPGE detector result in a charge being collected on the electrodes. The charge is converted into voltage-amplitude pulses by integrated cold and warm electronics. The detector is equipped with a charge-sensitive preamplifier with pulsed-reset feedback, helping to decrease the noise contribution of the feedback resistor~\cite{Barbeau:2007qi}. The preamplifier requires a periodical reset of the increasing baseline after the saturation of the dynamic range. This can be seen in a screenshot from the oscilloscope (see Figure~\ref{fig:pulses}).

The dynamic range of the preamplifier goes up to about 7~V, which is equivalent to about 2.3~MeV in the energy scale. A rather large signal immediately initiates the reset of the baseline and cannot be detected. The reset frequency depends on the sum of the detector leakage current and the count rate. For the \nuGeN detector inside the shielding at KNPP, the reset rate is about 5--10 Hz. The preamplifier has two similar amplitude outputs (OUT E and OUT E2), the inhibit output and the input for test signals. The inhibit output gives a logical signal at the time of the reset. The duration of the inhibit signal is set manually at 800 $\mu$s to exclude artificial signals generated by the reset. The signals from the output are shaped and amplified in order to obtain a positive signal suitable for the analog-to-digital converter (ADC). Parameters of each event (energy, timestamps) are evaluated by the multichannel analog-to-digital converter CAEN~VME~Realtime~ADC~V785N.

\begin{figure}[hbt]
  \begin{center}
    \includegraphics[width=1\linewidth]{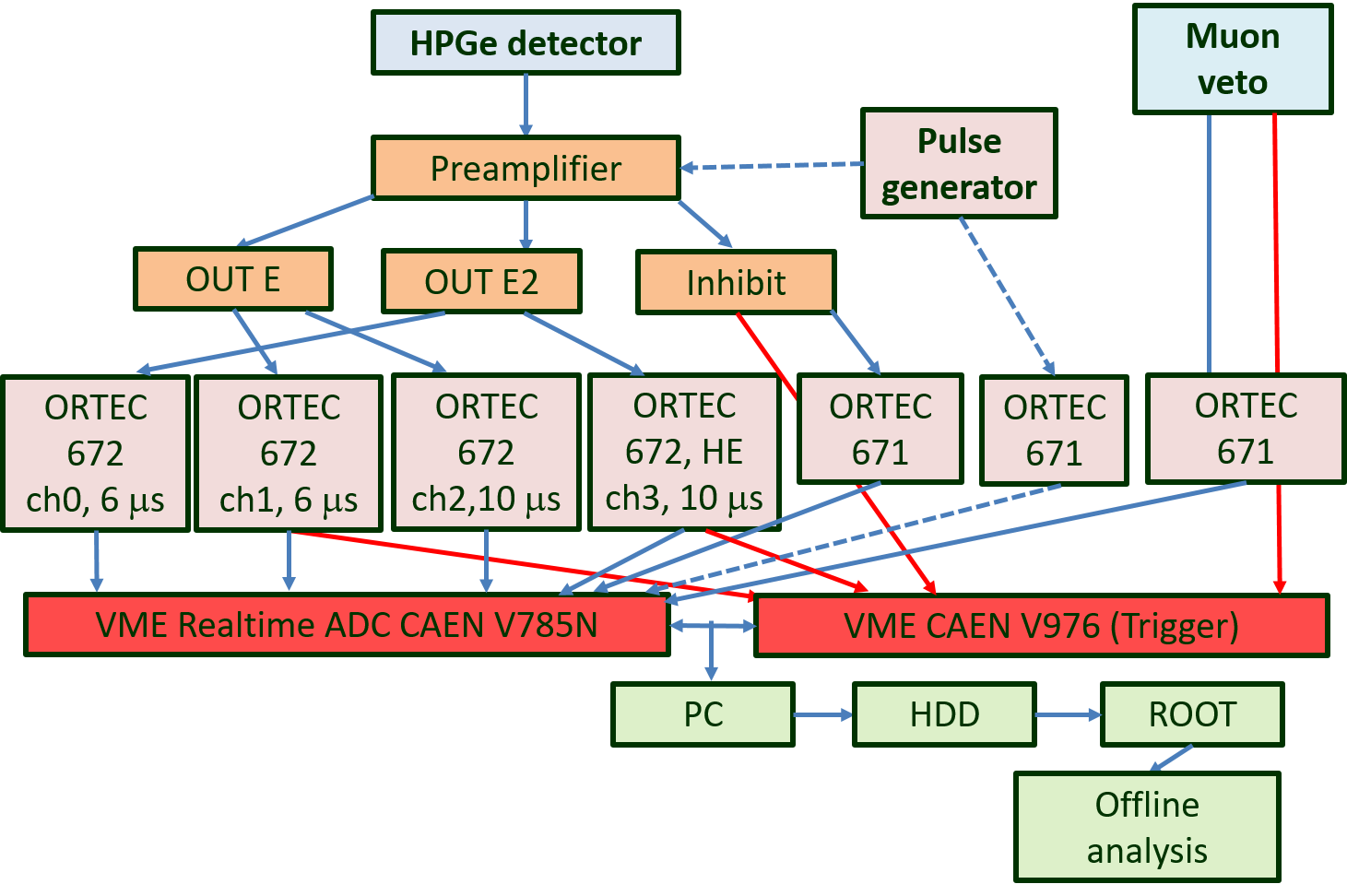}
    \caption{\label{fig:acq} Diagram of the acquisition system of the \nuGeN experiment.}
  \end{center}
\end{figure}

Each of the outputs of the preamplifier is connected to two ORTEC 672 spectroscopic amplifiers (four in total).
The output signals from each of the amplifiers are processed by ADC, giving four reconstructed energies that correspond to channels numbered from 0 to 3. Comparison and averaging of the signals reconstructed with different preamplifier outputs help to suppress the electronic noise. It also ensures the improvement in the energy resolution and decrease of the energy threshold.  Figure~\ref{fig:acq} shows a diagram of the components involved in the data acquisition. A wide energy range of up to 700~keV is measured with one of the ORTEC~672 amplifiers (labelled HE in Figure~\ref{fig:acq}). Timestamps from the muon veto system and inhibit signals are processed by the same ADC. The other channels are tuned up for measurements below $\sim$13~keV. The CAEN~VME~V976 trigger unit issues an acquisition command on input conditions corresponding to: 1)~low-energy HPGe signal; 2)~high-energy HPGe signal;~3) inhibit logical signal; 4)~muon veto.

The acquisition software of \nuGeN is based on the acquisition software previously designed for the DANSS experiment~\cite{Hons17},~\cite{Hons15}. It records real-time information about each event, including the amplitudes of all channels and the time of the event. Due to the KNPP safety restrictions, there is no internet access in the experimental hall, so the data are copied by shifters once per week for the offline analysis. The shifters also examine all equipment ensuring its stable and correct performance, and typically restart a new run every week or, if needed, more often. The offline analysis is performed later with the ROOT software~\cite{ROOT}. 

The cooling power of the CP5+ depends on the room temperature; therefore, an increase in temperature would increase the cooling power of the CP5+ and thus may lead to a higher level of noise in the detector. The experimental hall is equipped with air conditioners to decrease the temperature in the room and to provide stable ambient conditions. The temperature and humidity in the room are continuously recorded by two devices located in different places in the room. Typically, the temperature is about 22$^{\circ}$C, stabilized within $\pm$1$^{\circ}$C. The neutron background outside the shielding is controlled with a low-background neutron detector based on the CHM-57 counter~\cite{Rozov2010}, which was developed at JINR and used in a few experiments~\cite{Alekseev:2016llm},~\cite{Augier2023}.

The high-energy part of the spectrum is calibrated with a few-gram piece of a tungsten welding rod which contains about 2\% of $^{232}$Th. The energy calibration of the low-energy part of the spectrum is determined by means of the 10.37-keV cosmogenic line of $^{68,71}$Ge and artificial lines created by the pulse generators CAEN Mod. NTD6800D and ORTEC 419, which help to establish the calibration line slope. The shape of the generated pulses was rectangular, providing a response similar to the response of the physical pulse from the detector. The calibration of the low-energy part of the spectrum is verified by checking the position of the 1.3-keV cosmogenic line of EC decay of $^{68,71}$Ge. Typically, the calibration with the thorium source and pulse generator is performed monthly or after any change of the experimental conditions.

\section{Data selection and noise discrimination}

\begin{figure}[hbt]
  \begin{center}
    \includegraphics[width=1\linewidth]{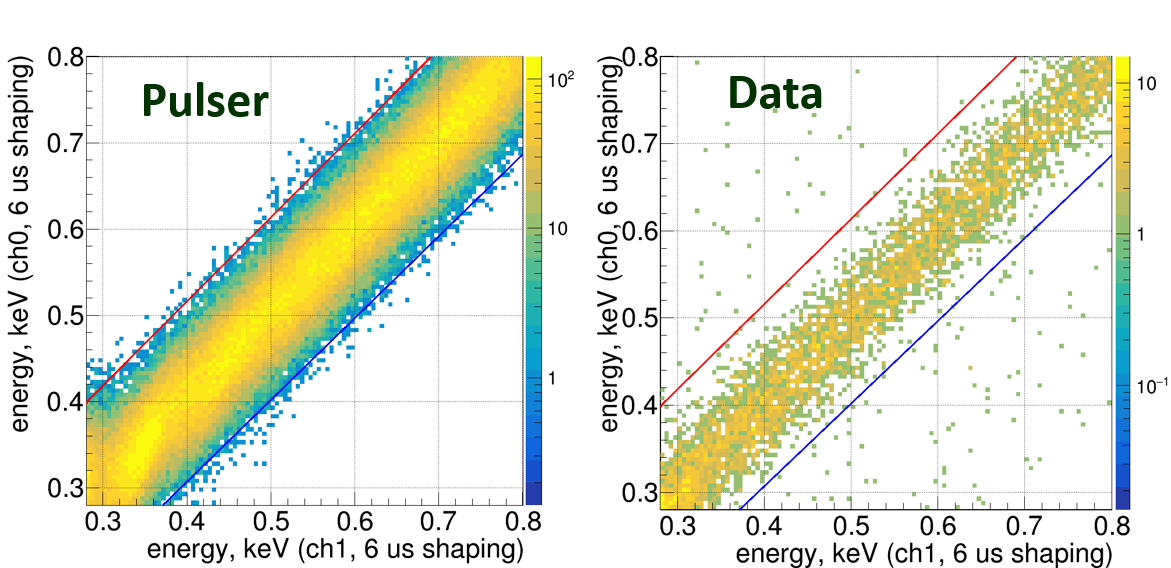}
    \caption{\label{fig:en1en0} Two-dimensional energy histograms from two identical channels for events from pulse generator (left) and for background events (right). Solid lines are graphical cuts to remove noise events.}
\end{center}
\end{figure}

The goals of the experiment require the best possible energy resolution and a low energy threshold connected to it. During preliminary measurements, it has been found that the lowest threshold and the best resolution are achieved with the 6-$\mu$s shaping time. The energy resolution obtained with the pulse generator is 101.6$\pm$0.5 eV (FWHM). In order to decrease the influence of the noise appearing in the electronic chain, the resulting energy of the signal is the combination of two reconstructed energies from different outputs of preamplifier with the weights of 2/3 and 1/3 for the channels 1 and 0, respectively. It was found that such averaging of the reconstructed energies reduces noise events and provides the lower energy threshold. Besides averaging, one can check whether the energy reconstructions in two different channels give similar results. The energy reconstruction differences indicate an impact of the electronic tract noise which could provide artificial events. Figure~\ref{fig:en1en0} shows the two dimensional histograms of the events reconstructed using different preamplifier chains. As shown in Figure~\ref{fig:en1en0}, the events from the pulse generator are located mostly diagonally, similar to the physical events. Using graphical cuts, which exclude events outside the diagonal area, it is possible to mitigate nonphysical noise events generated in the electronic chain.

\begin{figure}
  \begin{center}
    \includegraphics[width=1\linewidth]{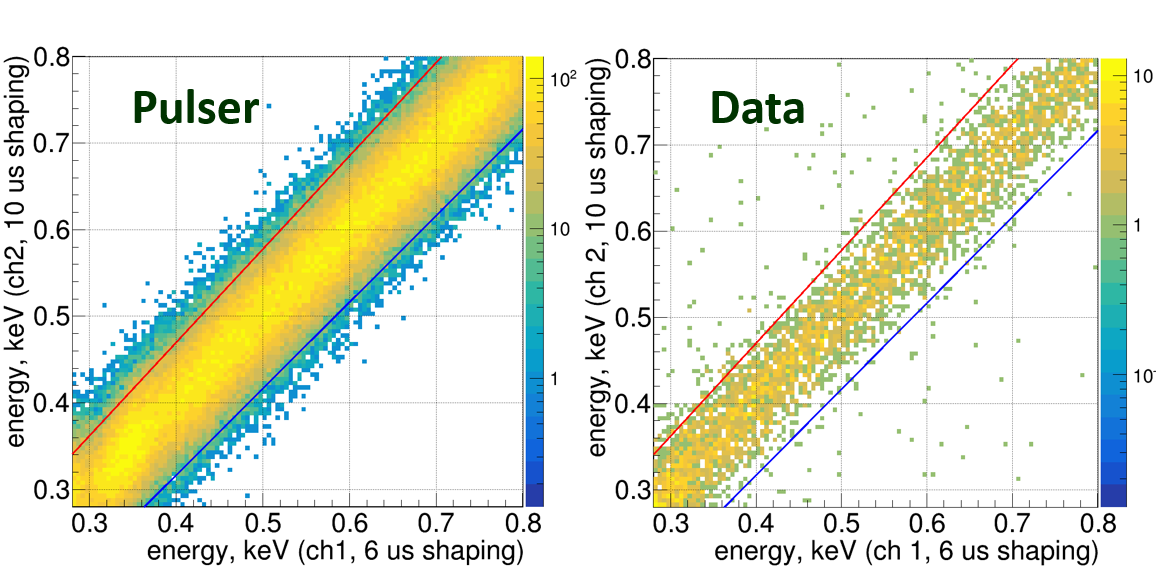}
    \caption{\label{fig:en1en2} Two-dimensional energy histograms from two channels with the different shaping times for events from the pulse generator (left) and for background events (right). Solid lines are graphical cuts to remove noise events.}
\end{center}
\end{figure}

In addition, the comparison of the signals obtained with different shaping times allows the efficient noise discrimination~\cite{Mor92},~\cite{Ste02}. We use the 6-$\mu$s and 10-$\mu$s shaping times to compare the signals. Similar to Figure~\ref{fig:en1en0}, graphical cuts allow suppressing the events with the noise origin due to their difference in energy reconstruction with different shaping times. Discrimination parameters of graphical cuts are shown in Figure~\ref{fig:en1en2}.

\begin{figure}[hbt]
  \begin{center}
    \includegraphics[width=1\linewidth]{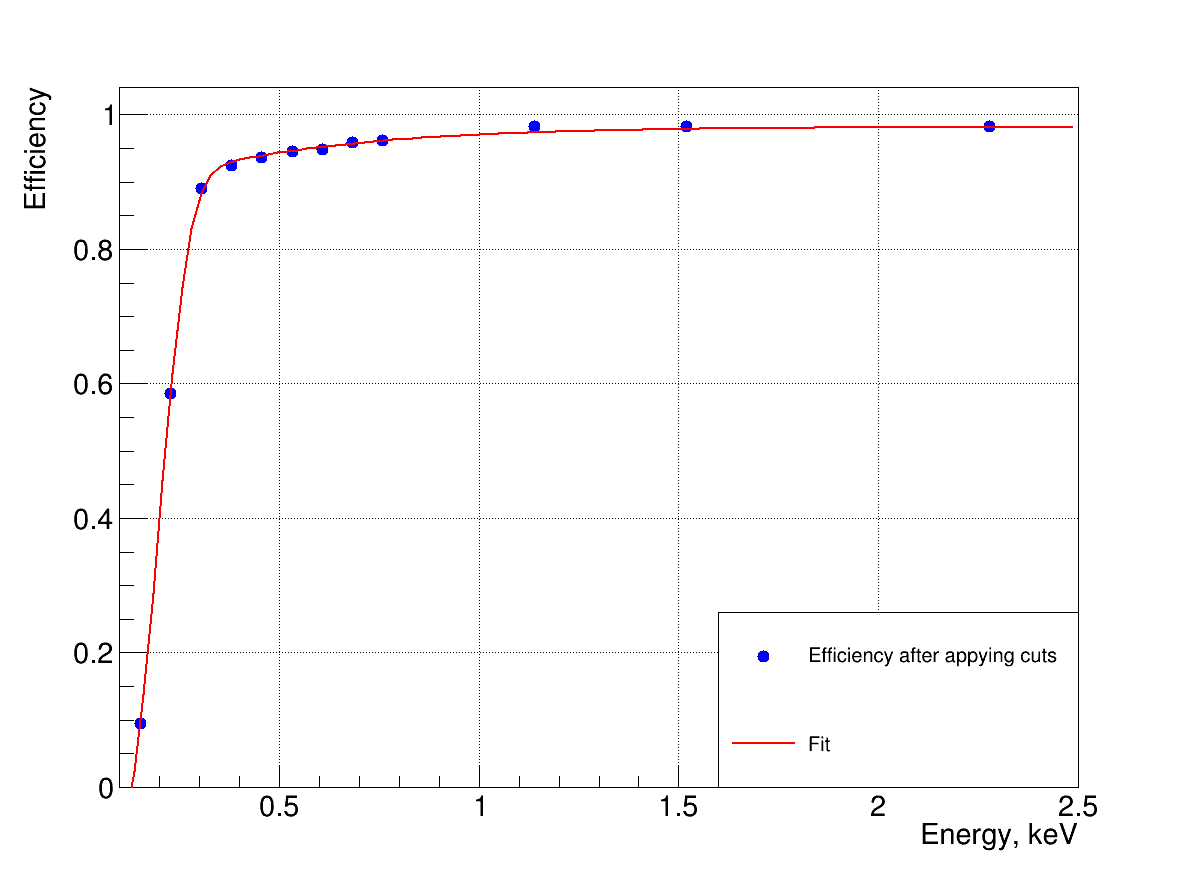}
    \caption{\label{fig:eff} Efficiency of signal detection measured with the pulse generator.}
\end{center}
\end{figure}

The efficiency of these cuts has been calculated by using measurements from the pulse generator. By comparing the observed intensity of the peaks in the energy spectrum to the rate of pulses from the generator, it is possible to determine the efficiency of graphical cuts and the trigger. The results of these investigations are shown in Figure~\ref{fig:eff}. It has been found that the detection efficiency is higher than 90\% for signals above 0.3 keV. To fit the efficiency distribution, we use the following function $F$:
\begin{equation}
F = \frac{1}{2} \Big( 1+ erf \big(\frac{E-a}{b} \big) \Big) + \Big(c-exp\big(d-\frac{E}{f}\big)\Big),
\end{equation}
where $E$ -- the energy, and $a,b,c,d,f$ -- the free parameters. The parameter values maximizing the likelihood of the fit are $a = 0.2006\pm0.0022$ , $b = 0.0821\pm0.0016$, $c = -0.0177\pm0.0033$, $d = -2.07\pm0.21$, $f = 0.42\pm0.09$. The obtained spectra are corrected according to the evaluated efficiency function. 

\begin{figure}[hbt]
  \begin{center}
    \includegraphics[width=1\linewidth]{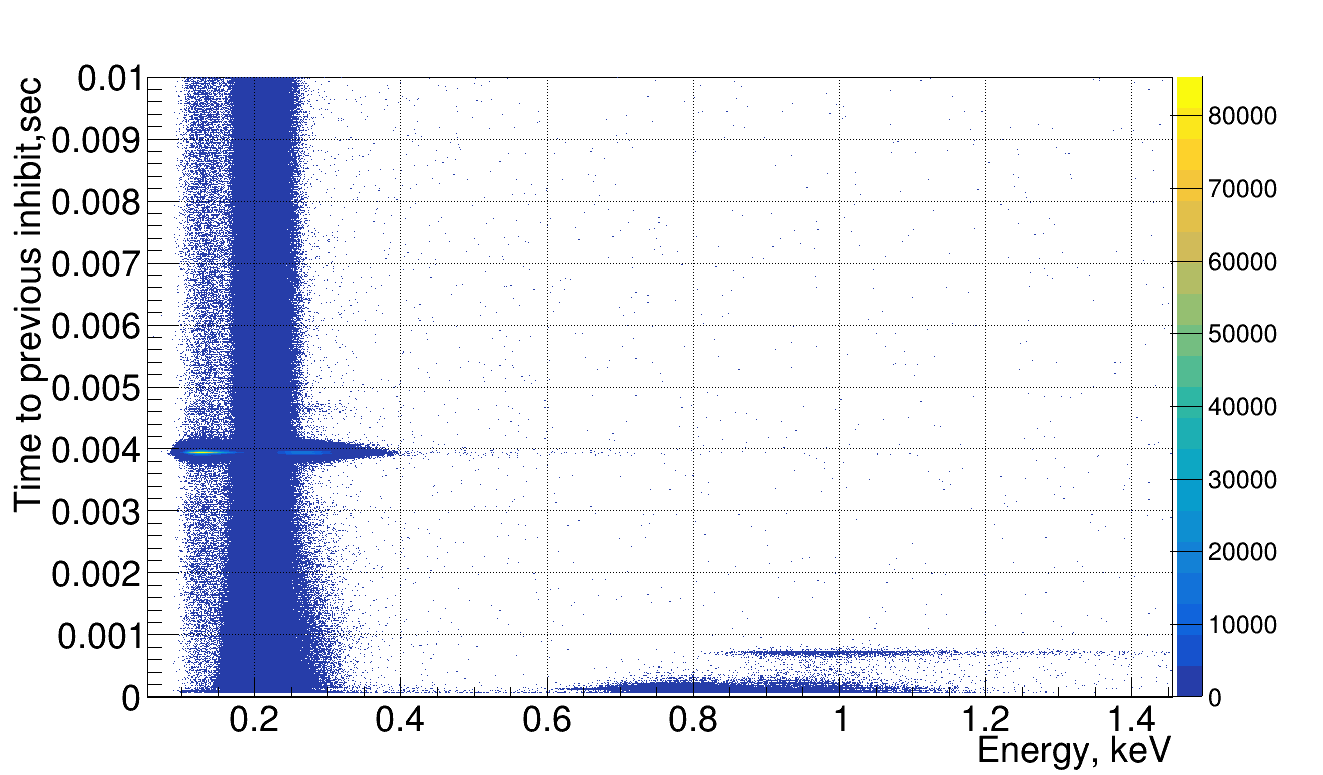}
    \caption{\label{fig:inh_cut_vs_energy} Distribution of time to the previous inhibit signals versus energy demonstrates events connected to the reset.}
\end{center}
\end{figure}

The reset of the baseline produces small afterpulses, which can also be interpreted as physical signals. Nonphysical events induced by the reset are clearly visible in Figure~\ref{fig:inh_cut_vs_energy}, which demonstrates the correlation between the energy of the event and the delay after the inhibit signal. To exclude these artificial events, the time period of 4.8~ms after each of the resets is not considered in the analysis.

The investigation of the time difference between consecutive events shows that there is another nonphysical population of signals, probably due to the reset and microphonic noises. These noise events can be mitigated by excluding any of the consecutive events within 150 $\mu$s. 

\begin{figure}[tbh]
  \begin{center}
    \includegraphics[width=1\linewidth]{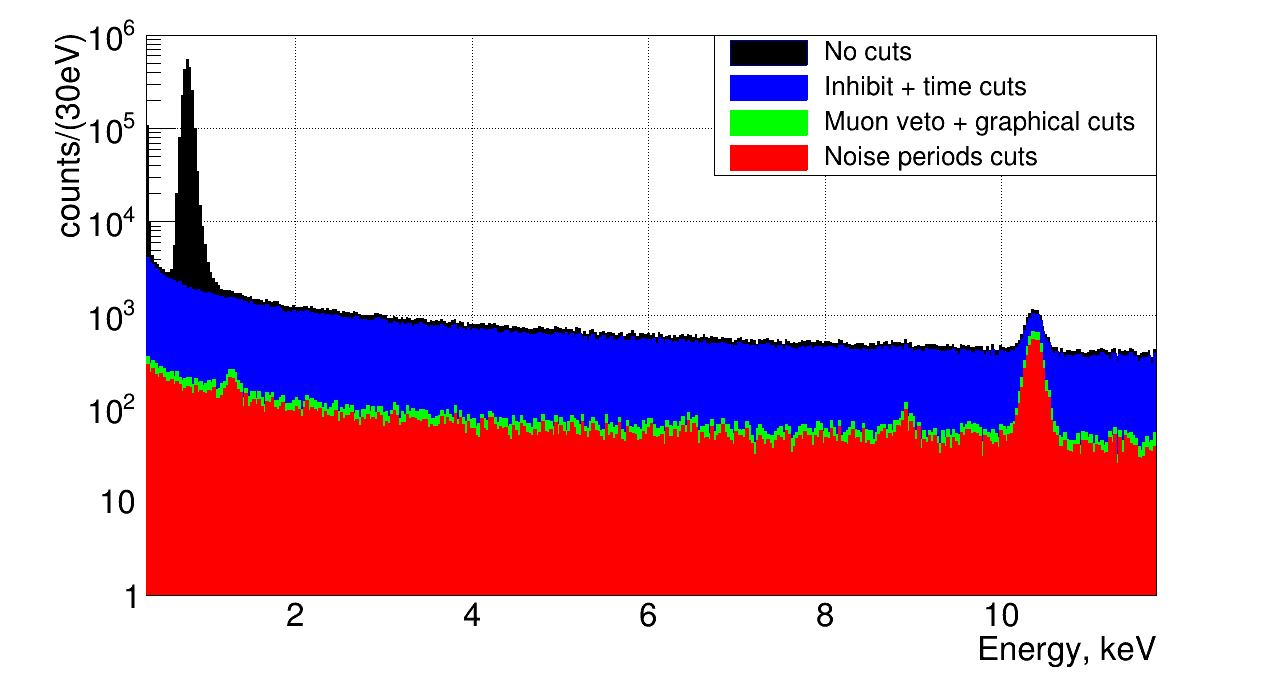}
    \caption{\label{fig:espectrum} Energy spectrum before (black) and after application of several cuts: inhibit+time cuts (blue), further application of muon veto and graphical cuts (green), and after removing time intervals with the high noise level (red) (see section~\ref{sec:stability} for details). }
\end{center}
\end{figure}

The energy spectrum before and after application of the quality cuts and the requirement of anti-coincidence with the muon veto is shown in Figure~\ref{fig:espectrum}. The artificial peak from the reset at about 0.8~keV is removed after applying the inhibit cut. 
The muon and inhibit cuts remove some of the physical events due to the dead time that is introduced after the muon or inhibit signals. The corresponding correction of the obtained spectrum with help of the calculated dead time of 9.0\% was performed.  The efficiency of the consecutive time cut was also taken into account by comparing the count rates before and after consecutive time cuts. The corresponding dead time from the consecutive time cut was found to be about 0.1\%.

\section{Stability in time}
\label{sec:stability}
 The stability of data taking is an important factor that affects further data interpretation. The direct comparison between ON and OFF datasets is not justified for the search for \CEvNS if the changes in the background and noise count rates are comparable to those of the expected signal. In this section we consider the factors potentially leading to the undesirable variations and describe selections ensuring the stability within the CEvNS dataset of September 2022 -- May 2023 collected at a distance of 11.1 m from the reactor core.
\begin{figure}
  \begin{center}
    \includegraphics[width=1\linewidth]{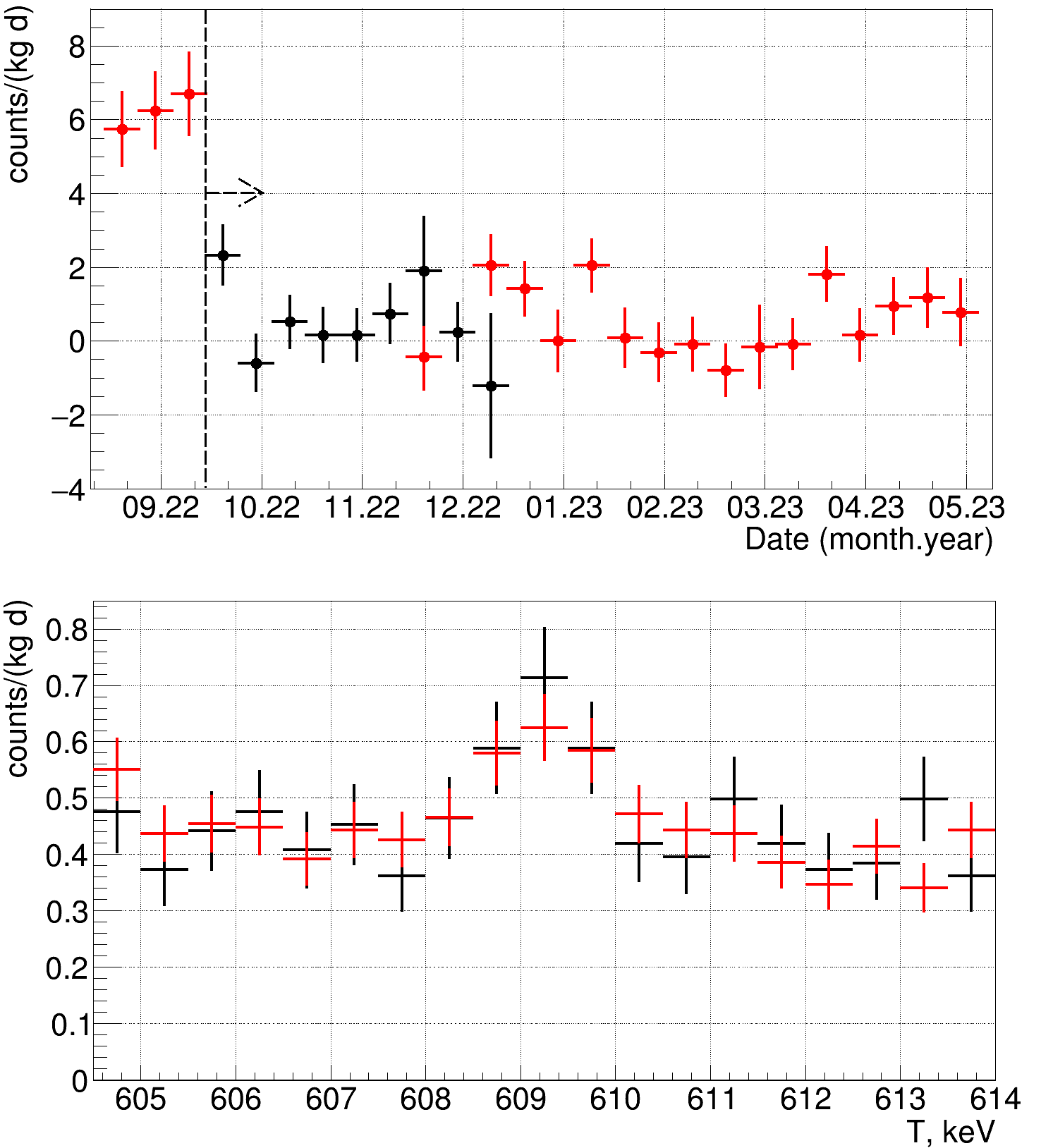}
    \caption{\label{fig:Rn} Top: count rate of the 609-keV $^{214}$Bi line vs. time for ON~(red) and OFF~(black) data, the dataset to the right from the dashed line is selected for further analysis. The ON and OFF periods overlap within the 10-day intervals (see also Fig.~\ref{fig:noise}). Bottom: accumulated count rate at 609~keV for ON~(red) and OFF~(black) data after the selection.}
\end{center}
\end{figure}

\paragraph{Radon-induced background.}
First, we consider the influence of $^{222}$Rn-related backgrounds on the \CEvNS region of interest~(ROI). The activity of radon inside the setup shielding was characterized on the basis of the 609-keV line from $^{214}$Bi. The stability of the corresponding rate in time and of accumulated counts is shown in the histograms in Figure~\ref{fig:Rn}. We exclude the part of the dataset acquired in September 2022 due to the increased radon activity. The remaining data suggest a stable radon concentration with no difference between ON and OFF datasets within the statistical uncertainty.

\paragraph{Cosmogenic background.}
The germanium detector was delivered to KNPP in 2019, and it was kept inside the shielding since November 2019. The intensity of the cosmogenic 10.37 keV line during reactor on and off is found to be 14.21$\pm$0.31 and 14.4$\pm$0.6 counts/(kg d), respectively. So no significant decrease of the gamma background is observed for the selected analysis period. 
\paragraph{Investigation of the noise  fluctuations.}
We make sure that the fluctuations of low-energy noise have only a negligible effect on the count rate in the ROI of CE{$\mathrm{\nu}$NS. For that purpose, we exclude the data taking periods when the changes of room temperature exceed the mean by the $\pm$1$^{\circ}$C. The variation of temperature can be connected with the noise level due to changes of the cryocooler power and corresponding mechanical vibrations. We also exclude the first 30 minutes of data after starting a new run and 10 minutes before the end of the run to avoid possible noise produced by on-site personnel. 
Noise events can significantly influence the energy spectrum. To exclude this, we have selected the energy threshold of about 0.29 keV above the region dominated by noise. After that, we check the stability of the count rate in the energy range from 0.25 to 0.28~keV, well below our analysis region and dominated by noise (see Figure~\ref{fig:noise}). 
\begin{figure}
  \begin{center}
    \includegraphics[width=1\linewidth]{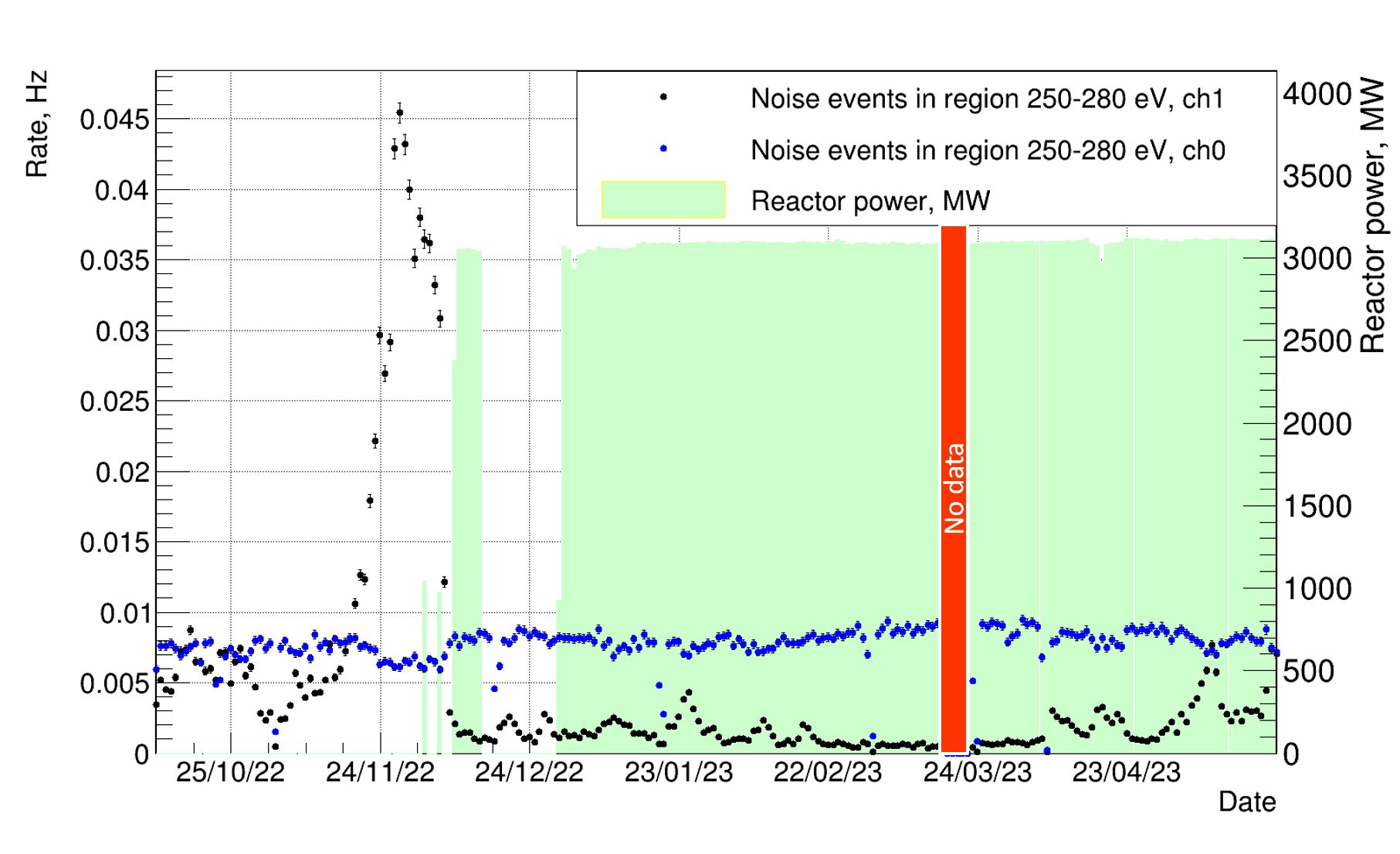}
    \caption{\label{fig:noise} Stability of the count rates observed in the noise energy region from 0.25 to 0.28~keV and reactor thermal power in time.}
\end{center}
\end{figure}
One of two channels (Channel 1 -- black dots) shows a significant time dependence. Although we do not observe any significant changes in the rates for the ROI, the time periods with the count rates~$>$~0.01 in the Channel 1 are excluded from the further analysis to avoid a possible influence of the noise on the spectrum.
\paragraph{Verification of the count rate stability.}
After exclusion of noisy time periods, we verify that residual noise fluctuations do not affect the ROI count rate by examination of its stability at the threshold. In particular, we fit the dependence of the count rate on time to the constant and check the $p$-values
\footnote{Probability of the $\chi^2$-score to be larger than that observed in the fit of the data if the tested hypothesis is true}
for the $\chi^2$-score for ON and OFF data separately. It is confirmed that the count rate becomes sufficiently stable for the threshold of 0.29~keV. The $p$-values of 94\% and 9\% were derived for ON and OFF data, respectively, in the energy interval from~0.29~to~0.31~keV. So, no statistically significant deviation from the constant count rate has been observed. We illustrate the stability achieved using the above selections by the count rate observed in the full \CEvNS ROI from~0.29~to~0.4~keV. Its dependence on time agrees with the constant for ON and OFF datasets analyzed separately, see Figure~\ref{fig:stab_roi}. The corresponding $p$-values are 82\% and 10\% for ON and OFF data.
\begin{figure}
  \begin{center}
    \includegraphics[width=1\linewidth]{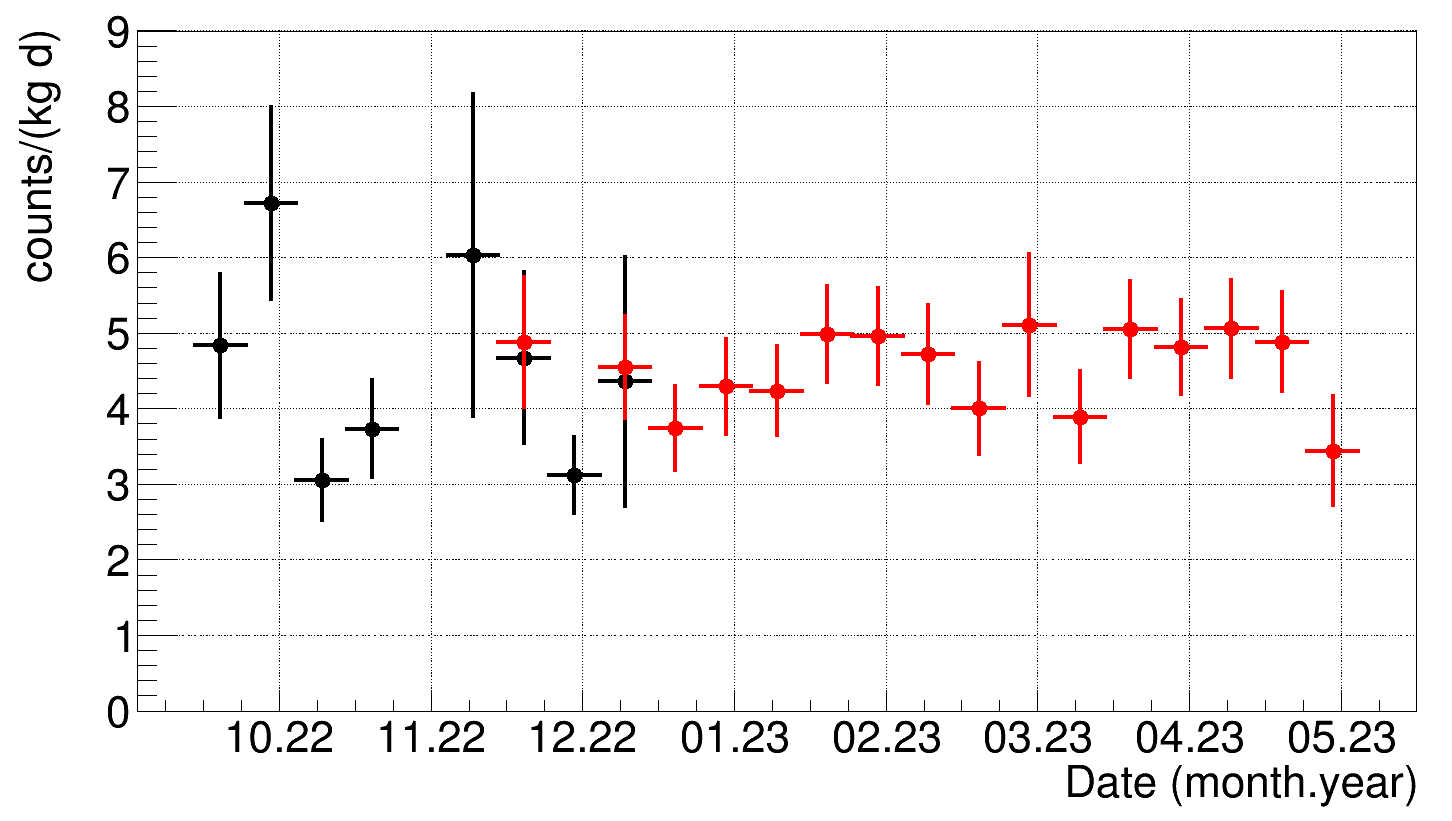}
    \caption{\label{fig:stab_roi} Stability of the count rate in \CEvNS ROI from 0.29 to 0.4~keV for ON (red) and OFF (black) periods. The energy-dependent selection efficiency constant is not corrected in time (see Fig.~\ref{fig:eff}).}
\end{center}
\end{figure}
After all selections, which ensure stability 194.5 and 54.6 kg$\cdot$days remain in the analysis for the reactor-ON and reactor-OFF periods, respectively.

\section{\label{sec:expect}Expected \CEvNS count rate}

The calculation of the expected spectrum of nuclear recoils from \CEvNS takes into account the antineutrino energy distribution and the parameters of the detector. The information about the isotope fraction and thermal power of the reactor was provided by KNPP personnel. The average fission fractions of the main isotopes of $^{235}$U, $^{238}$U, $^{239}$Pu, $^{241}$Pu are 0.642, 0.070, 0.246, and 0.042, respectively, for the analyzed period of the reactor-ON data. The calculation~\cite{Kop04} based on this fuel composition predicts 204.7~MeV of thermal energy per fission. The average thermal power evaluated for the reactor-ON dataset is 3081~MW. The reactor antineutrino energy spectra of up to 11~MeV were calculated using the summation model~\cite{Est19} based on the parameters described above.
This antineutrino spectrum was used to calculate the \CEvNS nuclear recoil energy distributions for each germanium isotope. The recoil spectra of five stable isotopes result in a summation spectrum.

The \CEvNS nuclear recoil spectrum must be modified to take into account quenching of ionization signals from nuclear recoils. This quenching is usually described by the Lindhard model~\cite{Lin63}. However, there is a significant discrepancy between the recent measurements at low nuclear recoil energies~\cite{Collar21, Bon22, COH_thesis, Kavner:2024xxd} affecting the predicted strength of the \CEvNS signal (see Figure~\ref{fig:quenching}). In this work, we consider three models for the quenching factor~(QF) dependence on energy. The first one is the Lindhard model ($k=0.162$), suggested by the measurements from CONUS~\cite{Bon22} (further referred to as ``$C$''). The decreasing trend of QF with the decrease of nuclear recoil energy is supported by ref.~\cite{COH_thesis}, although with somewhat lower QF values. The second model is suggested by the ``iron filter'' data of the Dresden-II experiment~\cite{Collar21}. This model (further referred to as~``$D1$'') assumes a linear fit of corresponding data below 1.35~keV$_{nr}$ combined with the Lindhard model~($k=0.157$) above that. The increase of QF with the decrease of energy is supported by ref.~\cite{Kavner:2024xxd}, which provides data for 254~keV nuclear recoils only. The latter demonstrate a QF quite lower than the one expected from the linear fit of~$D1$ (25\% vs. 38\% respectively). Finally, we consider the intermediate QF based on ``photo-neutron'' data of ref.~\cite{Collar21}. This model is further referred to as~``$D2$'' and is represented by the curve from the supplemental materials to ref.~\cite{Colaresi:2022obx}. The scenarios we consider illustrate the spread of existing experimental data. The first one~($C$) leads to a lower reactor \CEvNS count rate, while the second~($D1$) and the third~($D2$) demonstrate a more optimistic \CEvNS prediction. The resulting nuclear recoil spectra expected from \CEvNS in the \nuGeN detector for these QF models are shown in Figure~\ref{fig:cevns_gen}.
\begin{figure}[hbt]
  \begin{center}
    \includegraphics[width=1\linewidth]{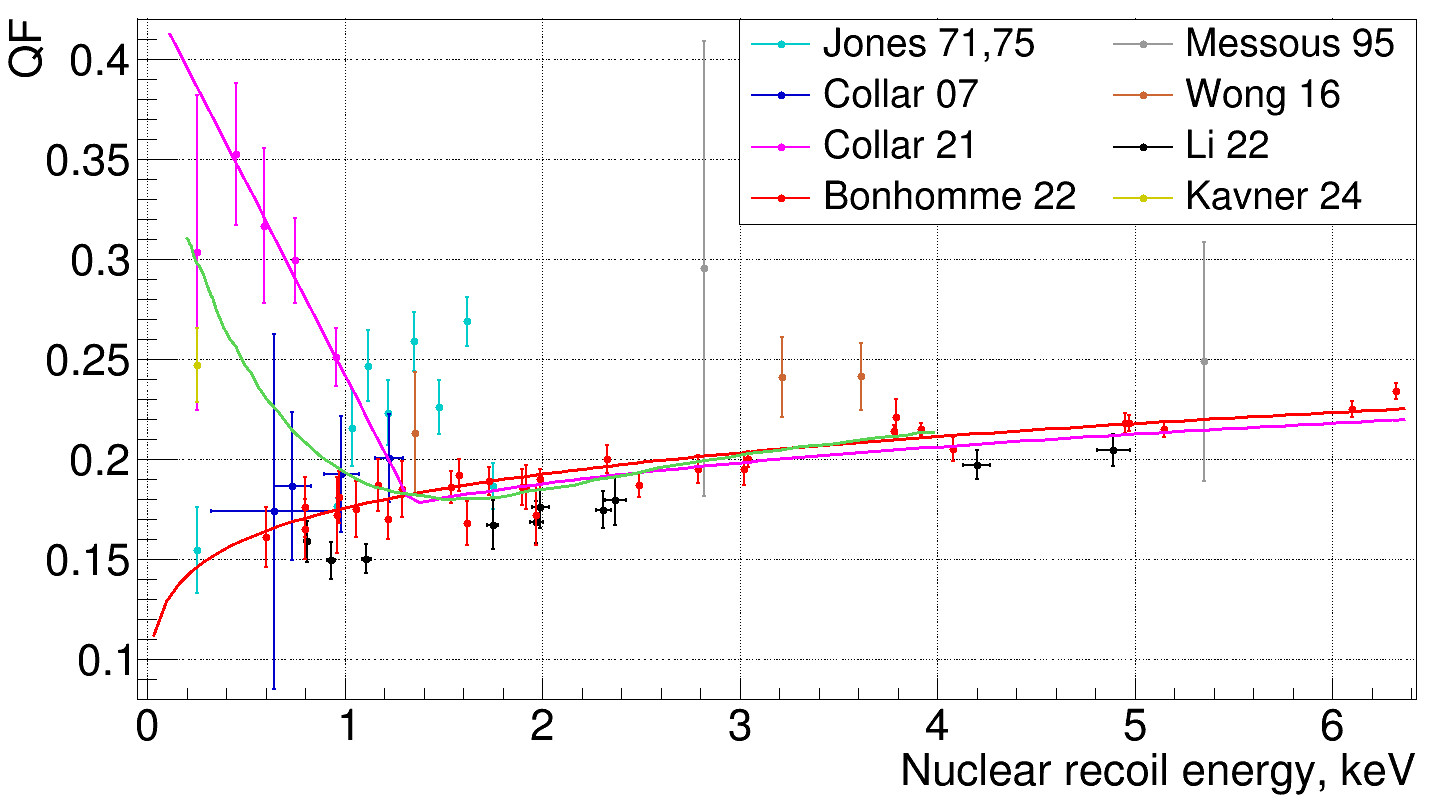}
    \caption{\label{fig:quenching} Measurements of germanium nuclear recoil QF~\cite{Jones:1971ya, Jones:1975zze, Mes95, Barbeau:2007qi, TEXONO:2014eky, Sch16, Collar21, Bon22, COH_thesis, Kavner:2024xxd}, the photo-neutron data from ref.~\cite{Collar21} are not shown for clarity, but are represented by one of the solid lines}. Solid lines: $C$~(red), $D1$~(magenta) and $D2$~(green) models, see text for details.
\end{center}
\end{figure}
\begin{figure}[hbt]
  \begin{center}
    \includegraphics[width=1\linewidth]{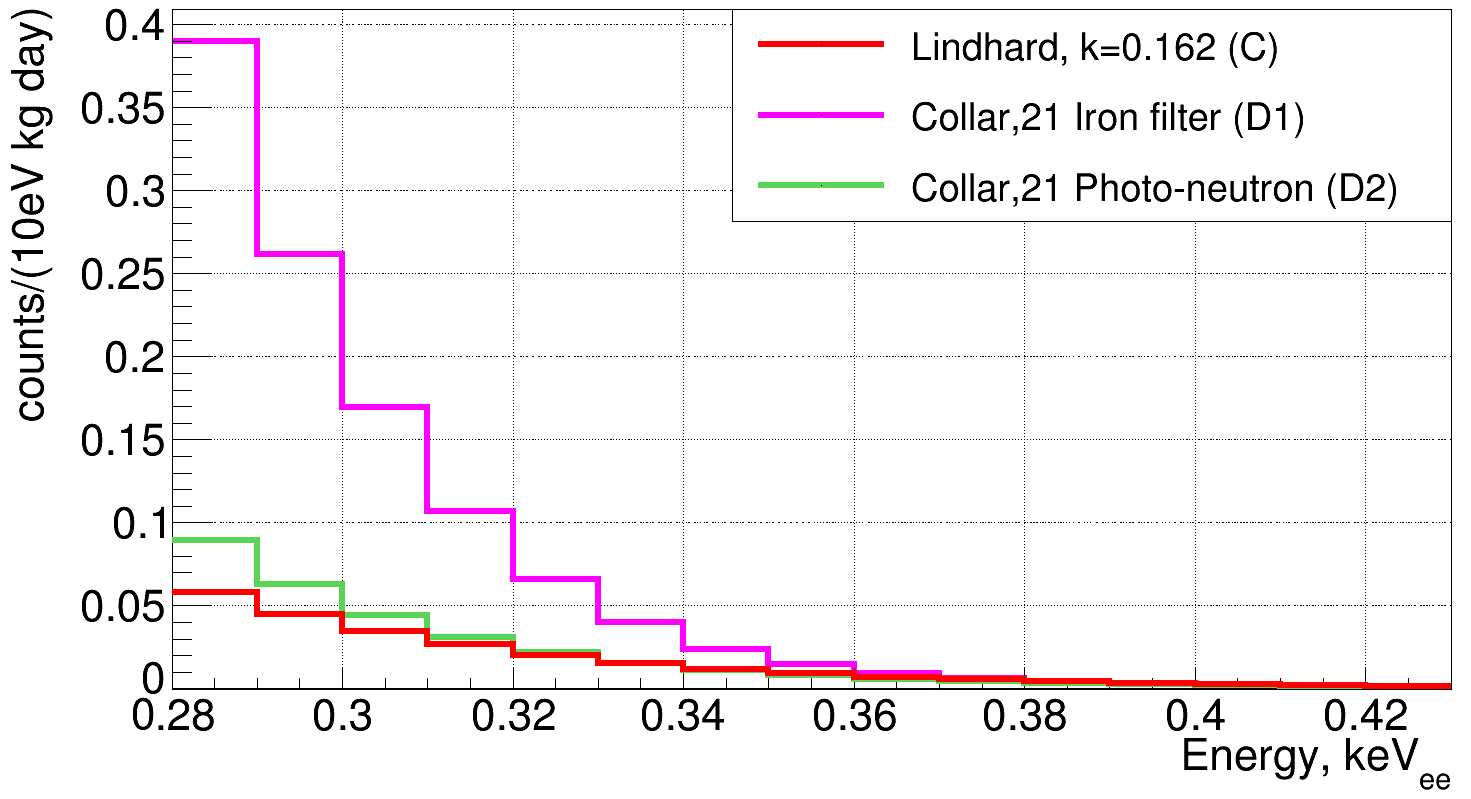}
    \caption{\label{fig:cevns_gen} Expected count rates and energy spectra of \CEvNS in the \nuGeN setup.}
  \end{center}
\end{figure}

\section{Results}
\paragraph{Statistical analysis of the residual spectrum.}
The experimental energy deposition spectra acquired both during reactor-ON and reactor-OFF periods are shown in Figure~\ref{fig:onoff}. To obtain these spectra, no scaling factors, besides the correction to the selection efficiency and live time, were used.

%
\begin{figure}[hbt]
  \begin{center}
    \includegraphics[width=1\linewidth]{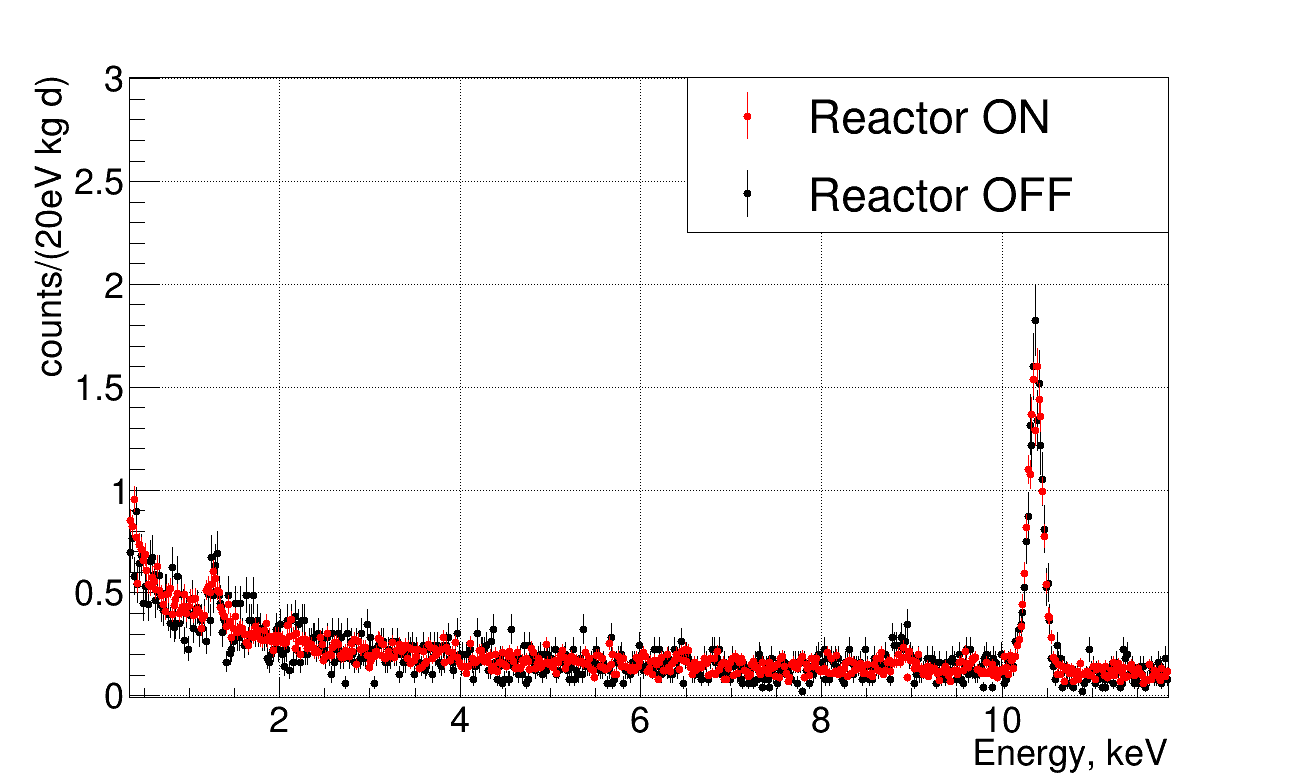}
    \caption{\label{fig:onoff} Comparison of the spectra acquired with reactor ON (red) and OFF (black).}
\end{center}
\end{figure}

We evaluate the constraints on the \CEvNS amplitude on the basis of the statistical analysis of the residual ON$-$OFF spectrum in ROI from 0.29 to 0.40~keV. The lower boundary of ROI was selected to exclude possible influence of the noise, and the higher boundary was selected from the analysis of the simulated spectra to maximize our sensitivity to \CEvNS for the current background level. The number of counts in each bin for both ON and OFF spectra is enough for the uncertainty of the residual spectrum to be a Gaussian one. We fit the residual spectrum using the \CEvNS prediction shape minimizing the $\chi^2$ statistics. The only free parameter in the fit is the \CEvNS amplitude $A$ in units of ``times Standard Model prediction''~($\times$SM). 
The results of the fit for the \CEvNS predictions based on the $C$, $D1$ and $D2$ quenching models are shown in Figure~\ref{fig:best_fit}~(top), the nominal predictions are presented for comparison (bottom).
%
\begin{figure}[hbt]
  \begin{center}
    \includegraphics[width=1\linewidth]{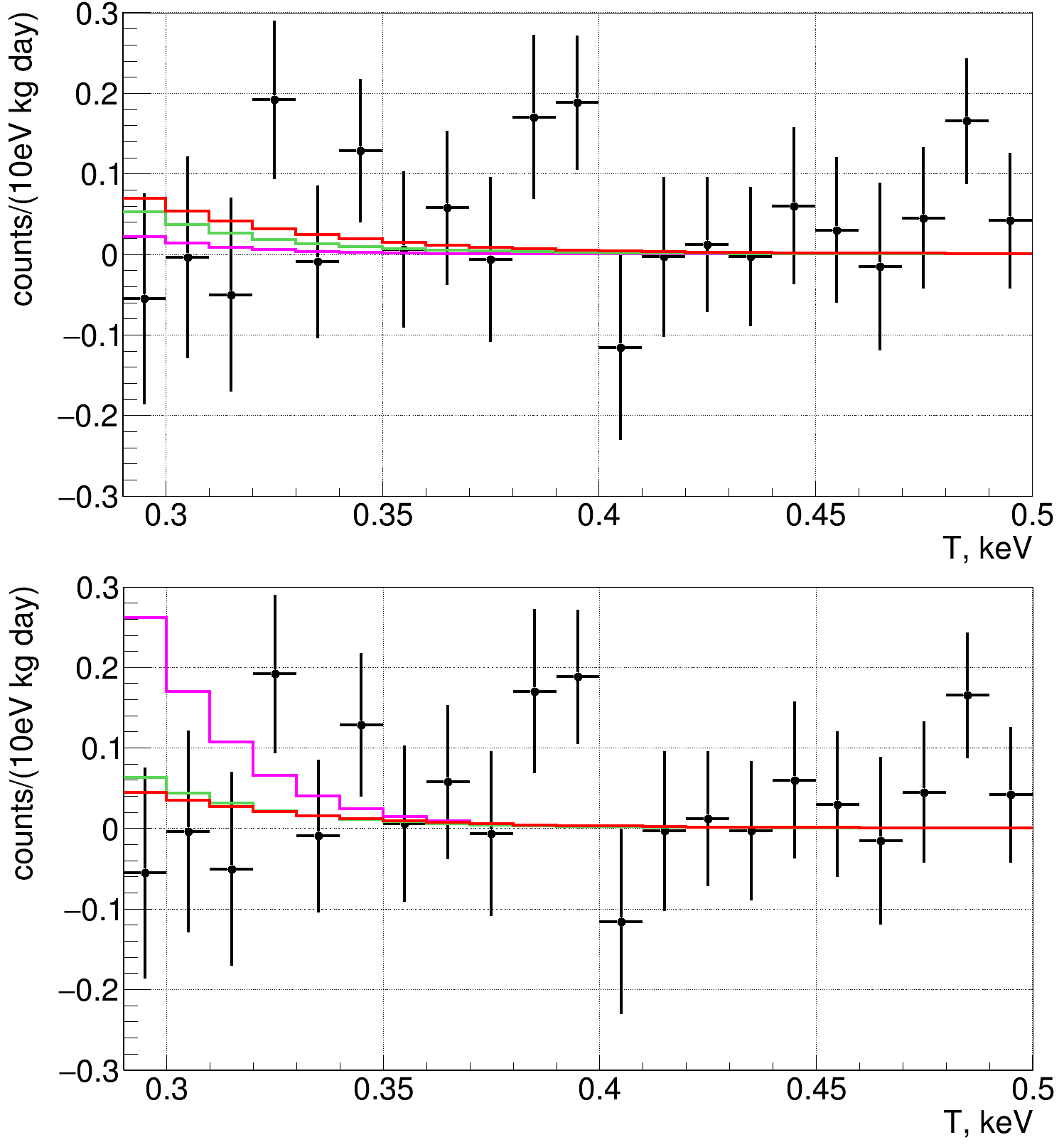}
    \caption{\label{fig:best_fit} The residual ON$-$OFF spectrum, the best signal shape fits (top) and the nominal signal predictions (bottom) for $C$~(red), $D1$~(magenta) and $D2$~(green)~QF.}
\end{center}
\end{figure}

Table~\ref{T_res} presents the best fit values $A_{best}$ and their statistical uncertainties $\sigma_{A}$, as well as the limits at a 90\% confidence level~(CL) on the \CEvNS amplitude. We also cite the value of the sensitivity, i.e. the expected median limit evaluated using the OFF data only. The $\Delta\chi^2$ profiles corresponding to the data fit are shown in Figure~\ref{fig:chi2}. It can be seen that the best fits under the assumption of $C$ and $D2$ QF do not contradict the Standard Model \CEvNS and the null hypothesis. The 90\%-CL upper limits on the \CEvNS amplitude are about 4.3/3.1 times larger than the Standard Model prediction respectively.

The fit for the most optimistic $D1$ QF model suggests the exclusion of the Standard Model \CEvNS rate at $2.5\sigma$. Such a result indicates either the presence of non-standard neutrino-quark interactions~(NSI) suppressing \CEvNS or a bias in the QF estimate. We confirm the earlier result of the CONUS experiment~\cite{Ackermann:2024kxo} excluding Standard Model \CEvNS for the $D1$ quenching scenario. Our result is also in tension with the reactor \CEvNS detection claim of the Dresden-II experiment~\cite{Colaresi:2022obx}. We note that the strength of this tension is challenged by one of the systematic uncertainties described below. For the completeness of discussion, the recent $3.9\sigma$ \CEvNS detection on Ge by COHERENT should be mentioned~\cite{Adamski:2024yqt}. The observed \CEvNS rate at the nuclear recoil energy range not affected by the QF discrepancy is about $2\sigma$ lower than the Standard Model prediction.
\begin{table}[hbt]
\caption{\label{T_res} Results of the statistical analysis of the residual spectrum, sensitivity ($S$) of the experiment and \CEvNS amplitude limits ($L$) at 90\% CL.}
\begin{center}
\begin{tabular}{ |p{0.6cm}|p{2.3cm}|p{2.0cm}|p{1.1cm}|p{1.1cm}| } 
 \hline
\centering QF & $A_{best}\pm \sigma_{A}$, $\times$SM & $\chi^2_{best}$~(ndf=10) & \centering S,$\times$SM & L,$\times$SM\\
\hline
\centering C & \centering $1.5 \pm 1.7$ & \centering 13.6 & \centering 3.8 & \centering 4.3 \tabularnewline
\centering D1 & \centering $0.1 \pm 0.4$ & \centering 14.4 & \centering 1.6 & \centering 0.7 \tabularnewline
\centering D2 & \centering $0.8 \pm 1.4$ & \centering 14.1 & \centering 3.3 & \centering 3.1 \tabularnewline

\hline
\end{tabular}
\end{center}
\end{table}
\begin{figure}[hbt]
  \begin{center}
    \includegraphics[width=1\linewidth]{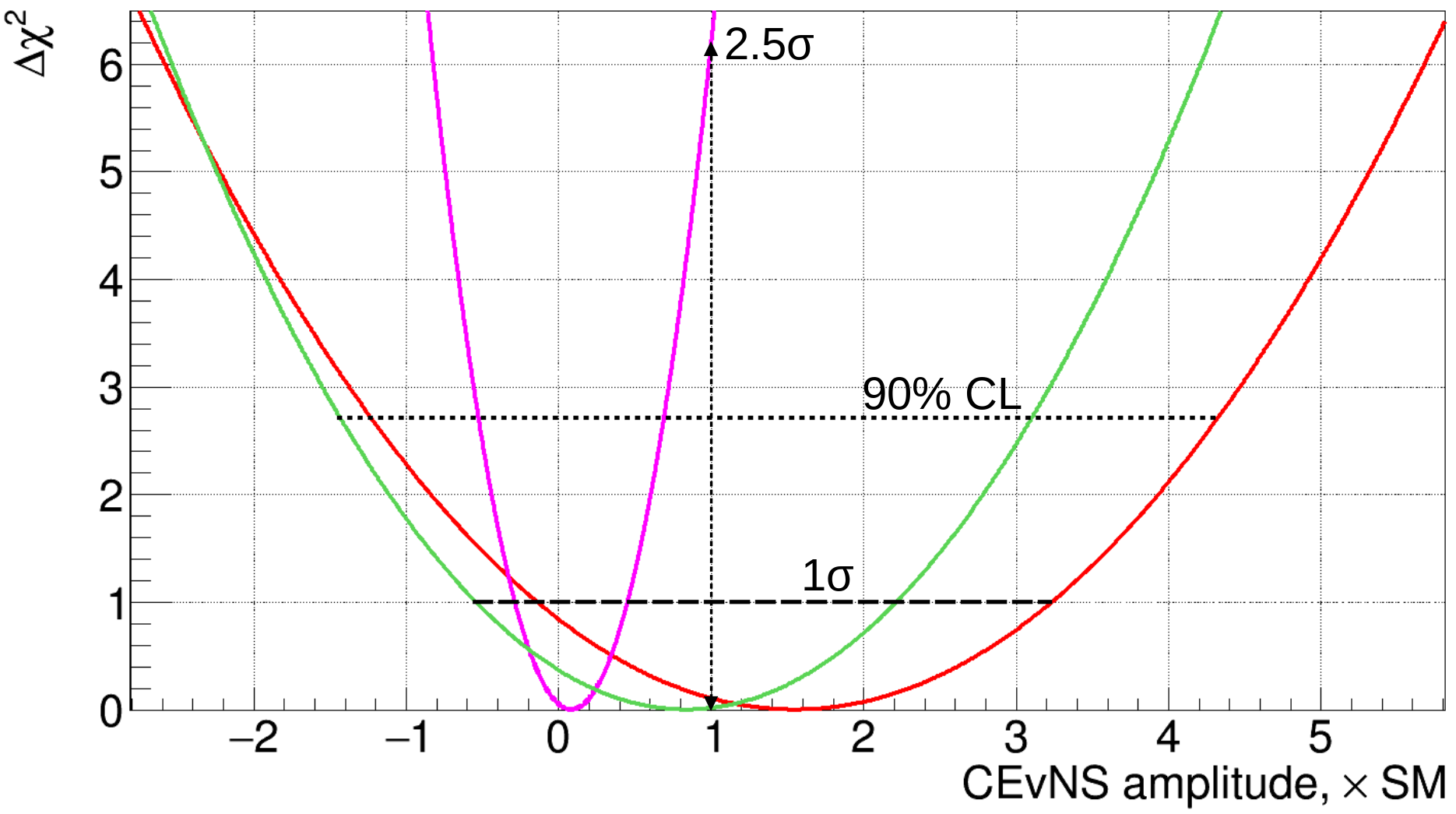}
    \caption{\label{fig:chi2} The profiles of $\Delta\chi^2$ statistics for the fit of the residual spectrum to the \CEvNS prediction shapes: $C$~(red), $D1$~(magenta) and $D2$~(green)~QF.}
\end{center}
\end{figure}
\paragraph{Systematic uncertainties.}
In what follows, we discuss the sources of systematic uncertainties and their impact on the result. The largest uncertainty is related to the nuclear recoil QF model. The resolution of the discrepancy between results of ref.~\cite{Bon22} and ref.~\cite{Collar21} requires more experimental data for the nuclear recoil energy below 1.3~keV. The measurements of CEvNS at reactors can provide another way to test QF, although the scenarios of NSI and an incorrect QF are hard to distinguish. Both \CEvNS amplitude and the QF-related parameter may be set free in the fit and tested with improvement of the energy threshold and new statistics in CONUS+~\cite{CONUS:2024lnu}, TEXONO~\cite{TEXONO:2024vfk}, Dresden-II, \nuGeN, and future germanium experiments. It is hard to say if this approach is feasible due to significant degeneracy between \CEvNS rate and QF. The Ge-based bolometers can test the \CEvNS amplitude independently from the ionization quenching if only the thermal signal is considered.

The second largest systematic uncertainty is associated with the energy scale calibration. To estimate this uncertainty, we use two modifications to the calibration procedure. The first ("global") one uses the entire statistics of the data without taking into account the small changes observed during regular calibrations, but with a higher statistics of the 10.37-keV calibration line. The second ("modified") one is performed to take into account a possible non-precise reconstruction of the energy scale at the low energy by using a slightly different slope of the calibration line leading to the shift of about 15 eV towards the high energies at ROI. Table~\ref{T_sys} demonstrates possible changes of the results due to the modification of the calibration energy scale. No significant effect on the final results in the cases of $C$ and $D2$~QF is observed. For the $D1$ scenario, the tension with the Standard Model \CEvNS decreased from $2.5\sigma$ to a modest $1.6\sigma$, below 90\%~CL.
\begin{table}[hbt]
\caption{\label{T_sys} The effect of the systematic uncertainty of the energy calibration scale on results for different QFs. The best fit values~($A_{best}$) and limits~(90\% CL) are in units of $\times$SM.}
\begin{center}
\begin{tabular}{ |p{1.4cm}|p{4.5cm}|p{2.5cm}| } 
 \hline
\centering Energy scale & \centering $A_{best}\pm \sigma_{A}$ (C/D1/D2) & Limit (C/D1/D2)\\
\hline
\centering Default & \centering $1.5 \pm 1.7$ / $0.1 \pm 0.4$ / $0.8\pm 1.4$ & \centering 4.3 / 0.7 / 3.1 \tabularnewline
\centering Global & \centering $1.8 \pm 1.7$ / $0.1 \pm 0.4$  / $1.0\pm 1.4$ & \centering 4.5 / 0.7 / 3.3 \tabularnewline
\centering Modified & \centering $1.2 \pm 2.4$ / $0.0 \pm 0.6$ / $0.6\pm 2.1$ & \centering 5.1 / 1.1 / 4.1 \tabularnewline
\hline
\end{tabular}
\end{center}
\end{table}

Another systematic uncertainty is associated with the reactor antineutrino energy spectrum. The \CEvNS count rate expected above the detector threshold is affected by the presence and magnitude of the high-energy part ($E_{\nu}>8$~MeV) of the antineutrino flux. In this work, we test two of the available models of spectra:  the Summation~Model~2018~\cite{Est19} (SM2018) and the model developed by authors from the Institute for Nuclear Research of the Russian Academy of Sciences in ref.~\cite{Vlasenko:2023eaf}~(INR RAS) verified using the data from the Double Chooz experiment~\cite{DoubleChooz:2019qbj}. Both models were implemented with average fission fractions of the $\nu$GeN exposition considered in this work. The results of these implementations are compared to each other and to the antineutrino energy spectrum measured by the Daya Bay experiment~(DB)~\cite{DayaBaySP1,DayaBaySP2HTR} in Figure~\ref{fig:spec_syst}. It can be seen that the count rate predicted  under the assumption of INR RAS is smaller than that for SM2018 by 10-15\% in the $\nu$GeN \CEvNS ROI. The DB-based estimate exceeds that for SM2018 by about 3\%. The effective fuel composition of DB is 56.4\%~($^{235}$U), 7.6\%~($^{238}$U), 30.4\%~($^{239}$Pu), and 5.6\%~($^{241}$Pu), slightly different from the average composition for the $\nu$GeN exposition. The estimates show that for the same fuel composition the discrepancy between SM2018 and DB increases up to about 5\%. So, non-negligible spread in the expected \CEvNS count rate should be considered as the systematic uncertainty of the results.

\begin{figure}[hbt]
  \begin{center}
    \includegraphics[width=1.0\linewidth]{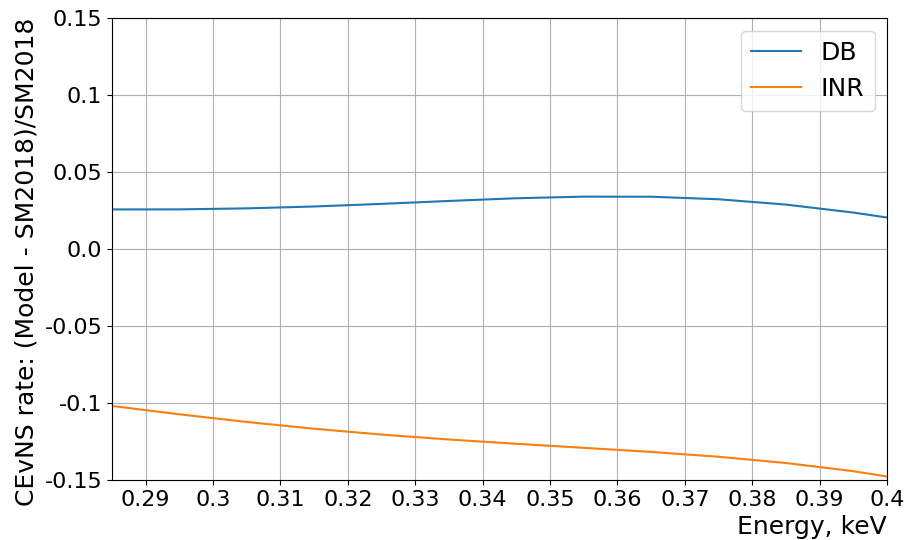}
    \caption{\label{fig:spec_syst} The effect of reactor antineutrino spectra models on the \CEvNS count rate estimates in the ROI of $\nu$GeN.}
\end{center}
\end{figure}

\section{Projected sensitivity}

To date, \CEvNS has been observed at the SNS using CsI[Na] scintillator~\cite{COHERENT:2021xmm}, liquid argon~\cite{COHERENT2}, and germanium~\cite{Adamski:2024yqt}. Recent results of experiments on the direct dark matter search~\cite{PandaX:2024muv,XENON:2024ijk} indicate the coherent scattering of solar boron neutrinos off xenon nuclei. The questionable claim of the \CEvNS detection at reactor~\cite{Colaresi:2022obx},~\cite{Ene23} and the most stringent upper limit (about 2 times above the SM for the $C$-model QF)~\cite{Ackermann:2024kxo} come from two germanium experiments, Dresden-II and CONUS
\footnote{During the review of this manuscript, the CONUS+ experiment reported a 3.7$\sigma$ observation of \CEvNS at a reactor in a preprint~\cite{Ackermann:2025obx}. The conclusions of ref.~\cite{Ackermann:2025obx} do not contradict results of this work.}, respectively. These two results, however, are in tension with each other under the assumption of the $D1$-model QF. Given this tension and a 2-$\sigma$ deviation from the SM of the result obtained in the COHERENT experiment~\cite{Adamski:2024yqt}, more data are required for the conclusive observation of \CEvNS at reactors, as well as the measurements of the \CEvNS cross-section on germanium. Considering this, the potential of the $\nu$GeN experiment is estimated below.

A typical reactor operation schedule at KNPP assumes 45 days OFF for every 16.5 months ON. Given this schedule, the statistical uncertainty of the residual count rate will be dominated by the OFF data. Using the background spectrum measured at KNPP during the reactor shut down, we extrapolate the sensitivity of $\nu$GeN for a larger expositions for $C$ and $D2$~QF. The significance of the expected null hypothesis rejection vs. data taking time is shown in Figure~\ref{fig:proj_sens}.
The ``time'' along horizontal axis in Figure~\ref{fig:proj_sens} corresponds to the accumulated OFF data (and assumes 11 times more ON than OFF at any moment). In this scenario, the 3-$\sigma$ null-rejection can be achieved only for the exposition longer than 27/19~years for $C/D2$ QF under the assumption of the 0.29-keV energy threshold. Another approach assumes using a background model instead of the OFF data~\cite{Bonet:2021wjw}. If the systematic uncertainty of the model based on the background decomposition is negligible comparing to the statistic uncertainty of the ON spectrum, then ``time'' in Figure~~\ref{fig:proj_sens} corresponds to ON exposition. In this case, the 3-$\sigma$ level can be achieved within few years.
\begin{figure}[hbt]
  \begin{center}
    \includegraphics[width=1\linewidth]{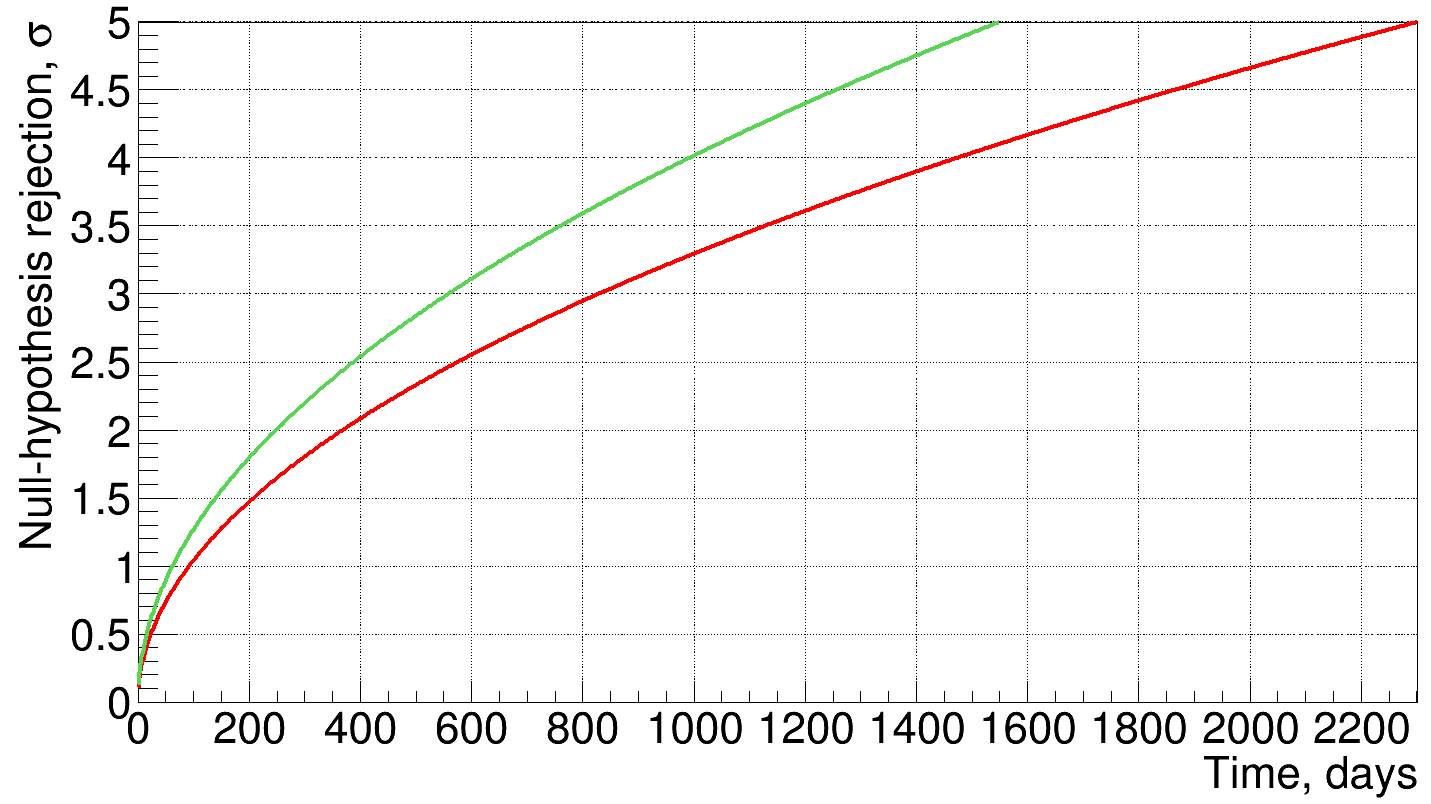}
    \caption{\label{fig:proj_sens} Projected sensitivity of $\nu$GeN based on the background count rate measured at KNPP for $C$~(red) and $D2$~(green) QF scenarios. The "time" on the X-axis corresponds to the OFF or ON statistics depending on the analysis strategy (see text for details).}
\end{center}
\end{figure}

Currently, we consider the decomposition of the $\nu$GeN background measured at KNPP, as well as its reduction by deployment of the NaI-based ``Compton Veto'' within the setup shielding. Additional laboratory tests are performed to improve the energy threshold of the HPGe detector by decreasing the power consumed by the cryocooler and extracting the waveforms from the detector for the offline pulse shape discrimination of surface events and noise~\cite{Bonet:2023kob}.

\section{Conclusion}

We considered the $\nu$GeN HPGe detector dataset, including 194.5~kg$\cdot$days of reactor operation and 54.6~kg$\cdot$days of reactor shut down, for search of the signal from coherent scattering of antineutrinos off germanium nuclei. The results of the fit of the signal prediction to the residual ON$-$OFF spectrum in the interval from 0.29 to 0.40~keV do not contradict both the null hypothesis and the Standard-Model \CEvNS under assumptions of $C$ and $D2$ QF scenarios. We evaluate the 90\%-CL upper limits of 4.3/3.1 times larger than the Standard-Model \CEvNS for $C$/$D2$ based on the statistical analysis of the data. Under the assumption of the most optimistic QF scenario $D1$, the data demonstrate the $2.5~\sigma$ tension with the Standard Model \CEvNS, indicating either the presence of non-standard interactions or the bias in the QF model. This latter discrepancy is softened by the systematic uncertainty of the energy scale calibration.
It is necessary to mention that our current analysis is based on the direct comparison of the count rates during reactor-ON and reactor-OFF regimes, which does not require the simulation of contributions of the background with additional assumptions of its shapes and level.

The projected sensitivity, estimated on the basis of the measured background count rate, shows the possibility of the 3-$\sigma$ \CEvNS detection within few years of stable data taking if the background model is used. The tests aimed at the background reduction and improvement of the energy threshold of the HPGe detector are carried out under laboratory conditions at the Joint Institute for Nuclear Research (JINR).

\section{Acknowledgements}

The authors are grateful to the KNPP directorate and staff for extensive support and practical help in performing measurements on the reactor site. The authors are thankful to Anton Lukyashin~(MEPhI, MIREA) and Valery Sinev~(INR RAS, MEPhI) for discussions about the reactor antineutrino energy spectra.

\providecommand{\noopsort}[1]{}\providecommand{\singleletter}[1]{#1}%

\end{document}